\documentclass[%
 reprint,
superscriptaddress,
nofootinbib,
 amsmath,amssymb,
 aps,prd,onecolumn
]{revtex4-2}
\usepackage{graphicx}
\usepackage{bm}
\usepackage{braket}
\usepackage{bbold}
\usepackage{amssymb}
\usepackage{amsmath,latexsym}
\usepackage{subfigure}
\usepackage{slashed}
\usepackage[T1]{fontenc}
\usepackage{multirow,array}
\usepackage{mathtools}
\usepackage{mathrsfs}
\usepackage{dsfont}
\usepackage{soul}
\usepackage[colorlinks=false,linktocpage=true]{hyperref}
\usepackage{hyperref}
\usepackage[utf8]{inputenc}
\usepackage[export]{adjustbox}
\usepackage{float}
\usepackage[makeroom]{cancel}
\usepackage{wrapfig}
\usepackage{ulem}
\usepackage{graphicx} 
\usepackage{amsmath}
\usepackage{physics}
\usepackage{color}
\usepackage{xcolor}
\usepackage{orcidlink}

\newcommand{\be}{\begin{equation}}
\newcommand{\ee}{\end{equation}}
\newcommand{\ba}{\begin{eqnarray}}
\newcommand{\ea}{\end{eqnarray}}

\begin{document}

\newcommand*{\Bogazici}{Department of Physics, Bogazici University, Istanbul, Turkey}  

\newcommand*{\UConn}{Department of Physics, University of Connecticut,
  Storrs, CT 06269, U.S.A.}
  
\newcommand*{\BNL}{Department of Physics, Brookhaven National Laboratory, Upton, NY 11973, U.S.A.}

\title{\boldmath
Unified genus-1 potential and a parametric  perturbative/nonperturbative relation}

\author{Atakan \c{C}avu\c{s}o\u{g}lu\,\orcidlink{0009-0005-5268-3266}}
\email{atakanc@sas.upenn.edu}
\altaffiliation{Present address: Department of Physics and Astronomy, University of Pennsylvania, Philadelphia, Pennsylvania 19104, USA.}
\affiliation{Department of Physics, Bo\u{g}azi\c{c}i University, Istanbul, Turkey}
\affiliation{Department of Electrical and Electronics Engineering, Bo\u{g}azi\c{c}i University, Istanbul, Turkey}

\author{Can Koz\c{c}az\,\orcidlink{0000-0002-7768-926X}}
\email{can.kozcaz@bogazici.edu.tr}
\affiliation{Department of Physics, Bo\u{g}azi\c{c}i University, Istanbul, Turkey}
\affiliation{Feza G\"{u}rsey Center for Physics and Mathematics, Bo\u{g}azi\c{c}i University, Istanbul, Turkey}
\affiliation{Niels Bohr Institute, Copenhagen University, Blegdamsvej 17, Copenhagen, 2100, Denmark}

\author{Kemal Tezgin\,\orcidlink{0000-0001-9492-9512}}
\email{kemaltezgin@vt.edu}
\altaffiliation{Present address: Department of Physics, Virginia Tech, Blacksburg, Virginia 24061, USA.}
\affiliation{Department of Physics, University of Connecticut, Storrs, Connecticut 06269, USA}
  
\begin{abstract}
\noindent
We study a parametric deformation of the unified genus-1 anharmonic potential and derive a parametric form of perturbative/nonperturbative (P/NP) relation, applicable across all parameter values. We explicitly demonstrate that the perturbative expansion around the perturbative saddle is sufficient to generate all the nonperturbative information in these systems. Our results confirm the known results in the literature, where the cubic and quartic anharmonic potentials are reproduced under extreme parameter values, and go beyond these known results by developing the nonperturbative function of real and complex instantons solely from perturbative data.
\end{abstract}

%
%


\maketitle

\section{Introduction}
\label{sec-1:Introduction}
\par{The spectrum of only a small number of quantum mechanical systems can be exactly determined. One of the most successful approaches to approximately determining the energy eigenvalues and the associated eigenstates has been the perturbation theory, which relies on the knowledge of an exactly solvable system, such that the system in question can be regarded as a small deformation of the known solvable system. In addition to the obvious limitations of perturbation theory due to its very definition, there are even more serious problems with this technique. First, the perturbative expansion generically has zero radius of convergence due to the factorial growth of the large-order terms in the series. In many applications, the perturbative series is cut short by ignoring all terms beyond a given order and considering only the corrections due to the low order contributions. Although this approach provides even experimentally verifiable results, it misses most of the physical information encoded in the perturbative expansion. As a remedy, one might be tempted to apply one of the celebrated re-summation techniques, such as the Borel re-summation, to assign a well-defined value to the divergent perturbative series at hand. However, such a re-summation by itself might not always be practical due to the possible ambiguities and maybe even worse because of the complex-valued corrections to physically observable quantities, such as the energy levels for stable potentials.}
\par{Another important shortcoming of the perturbation theory is its failure to capture the classically forbidden phenomenon of quantum tunneling. It is known that the energy spectrum of a system gets exponentially suppressed (nonanalytic) corrections due to tunneling, which we refer to as instantons. Since they are beyond the reach of the perturbation theory per se, we refer to their contributions as \textit{nonperturbative} corrections (we will soon argue that the perturbative and nonperturbative corrections are ultimately intertwined). Moreover, in the presence of more than one instanton, the correct spectrum can only be obtained if the perturbative and nonperturbative corrections are accounted for in a unified fashion, including the interactions between instantons. }

\par{The intricate structure of the energy levels can be recast in the form of a transseries, which not only takes into account the perturbative corrections but also the nonanalytic corrections due to the presence of instantons and their interactions. Generically, we are aiming to express the energy level using an expansion of the form \cite{Zinn-Justin:2004vcw, Zinn-Justin:2004qzw, Jentschura:2010zza}}
\begin{align}
E^{(N)}(\hbar)=\sum_{\pm}\sum_{n=0}^{\infty}\sum_{l=1}^{n-1}\sum_{m=0}^{\infty}c^{\pm}_{nlm}\frac{e^{-n\frac{S}{\hbar}}}{\hbar^{n(N+\frac{1}{2})}}\left(\ln\left[\mp\frac{2}{\hbar}\right]\right)^{l}\hbar^{m},
\end{align}
where the exponential monomials are due to the instantons, and the logarithmic monomials account for their interactions. The series not only includes the perturbative fluctuations around the perturbative vacuum but also is made of fluctuations around different numbers of instantons. The resurgent structure \cite{Ecalle:1981} is responsible for precise relations among the different coefficients $c^{\pm}_{nlm}$, such that they render an ambiguity-free and real expression for the energy levels after Borel resummation.
\par{Zinn-Justin and Jentschura proposed a generalization of the Bohr-Sommerfeld quantization condition to take into account the nonperturbative effects as well. The generalized quantization condition is expressed in terms of two functions $B(E,\hbar)$ and $A(E,\hbar)$, which respectively encode the perturbative and nonperturbative aspects of the system. More precisely, if the function $B(E,\hbar)$ is inverted and written as $E(B,\hbar)$, it becomes identical to the usual perturbative series after setting $B=N+\frac{1}{2}$. On the other hand, the function $A(E,\hbar)$ starts with the instanton action $S/ \hbar$ and includes the quantum fluctuations around an instanton. For instance, the generalized quantization condition for the double-well is given by \cite{Zinn-Justin:2004qzw, Zinn-Justin:2004vcw}}
\begin{align} \label{DoubleWellQuantizationCondition}
\frac{1}{\sqrt{2\pi}}\Gamma\left(\frac{1}{2}-B(E,\hbar)\right)\left(-\frac{2}{\hbar} \right)^{B(E,\hbar)}\exp\left[-\frac{A(E,\hbar)}{2} \right]=\pm i.
\end{align}
This equation can be recursively solved for the energy levels $E$, including the nonperturbative corrections to it. Recently, a nice derivation for this type of quantization conditions was offered in the context of the exact WKB approach as the necessary condition for the wave function to be normalizable \cite{Sueishi:2020rug, Sueishi:2021xti, Kamata:2021jrs}.
\par{Usually, these two functions are regarded as independent quantities since they describe completely different physics and need to be calculated separately. This view was dramatically changed when Dunne and \"{U}nsal  \cite{Dunne:2013ada} reinterpreted the explicit perturbative and nonperturbative relations for one-dimensional oscillator discovered earlier by \'{A}lvarez \cite{Alvarez2004} as a constructive approach to obtain the nonperturbative contributions encoded in $A(E,\hbar)$ from the perturbative information concealed in $B(E,\hbar)$,   }
\begin{align}
A(E,\hbar)=- \int \frac{d\hbar}{\hbar}\left( \frac{S}{\hbar} \frac{\partial E}{\partial B}+B\right),
\end{align}
which we call the P/NP relation\footnote{Also referred to as Dunne-\"{U}nsal relation in the literature.}. In other words, the knowledge of the instanton action and the perturbative expansion to any order is sufficient to obtain the quantum fluctuations around the instanton to the same order. It is important to emphasize, however, that the large-order coefficients of the perturbative series already contain the information about the instanton action, implying that perturbative expansion around the perturbative saddle encodes all the required details to derive information about the instanton function $A$. Supplemented with the quantization condition, the exact energy levels can be entirely obtained from the perturbative expansion around the perturbative saddle. 

The P/NP relation can be equivalently expressed as a differential equation,
\begin{align} \label{Dunne-Unsal}
\frac{\partial E(B,\hbar)}{\partial B}=-\frac{\hbar}{S}\left(B+\hbar\frac{\partial A(B,\hbar)}{\partial\hbar} \right).
\end{align}
One of our main results is the generalization of the P/NP formula to a two-parameter family of quartic potentials, interpolating between pure cubic and quartic potentials, for which the P/NP relation has already been confirmed \cite{Alvarez:2000, Alvarez:2002, Gahramanov:2015yxk}. Although our formula is in the spirit of the P/NP relation, in the sense that it allows us to construct the nonperturbative function $A(E,\hbar)$ from the perturbative information $B(E,\hbar)$, its functional form is different since it includes derivative terms with respect to two deformation parameters. 
\par{The paper is organized as follows: in the next section, we first review the Picard-Fuchs equation to compute the classical periods on the hyperelliptic curve associated with our deformed potential. Then, we focus on the exact WKB approach and the quantum corrections to the classical periods. After we elaborate on the precise relationship of the quantum corrected periods to the aforementioned functions $A(E,\hbar)$ and $B(E,\hbar)$, we state the parametric version of the P/NP relation, which is a generalization to a family of potentials interpolating between the cubic and quartic potentials. In section 3, we study the large-order behavior of the perturbation theory in the presence of either real or complex instantons. We demonstrate how the corrections improve the match between the large-order perturbative series and the early terms of instanton fluctuations, reminiscent of the more customary behavior from the resurgence theory. Finally, we conclude with a short review of our results.}
\par{Resurgence theory has been applied in various quantum mechanical systems \cite{Zinn-Justin:2004qzw, Zinn-Justin:2004vcw, Jentschura:2010zza, Alvarez:2000, Alvarez:2002, Dunne:2013ada, Dunne:2014bca, Basar:2013eka, Cherman:2014ofa, Behtash:2015loa, Behtash:2015zha, Gahramanov:2015yxk, Gahramanov:2016xjj, Basar:2015xna, Dunne:2016qix, Fujimori:2017osz, Fujimori:2018kqp, Fujimori:2022lng, Codesido:2017dns, Basar:2017hpr, Gorsky:2014lia, Misumi:2015dua, vanSpaendonck:2023znn}, and for reviews on the subject, we refer to \cite{Costin:2008, Dorigoni:2014hea, Marino:2015yie, Aniceto:2018bis}.}

\section{Picard-Fuchs equations and the P/NP relation} \label{Section:PFEquationAndPNP}

\begin{figure}
\begin{centering}
\includegraphics[width=5.6cm]{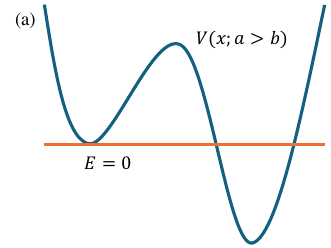} \hspace{2mm} \
\includegraphics[width=5.6cm]{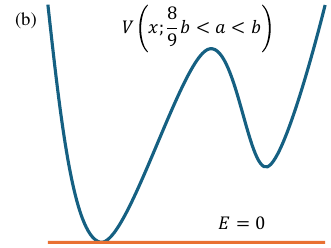} \hspace{2mm} \
\includegraphics[width=5.6cm]{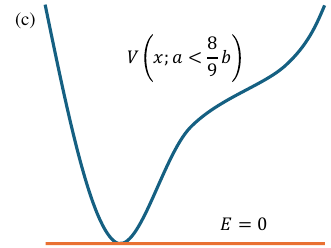}
\par\end{centering}
\caption{\label{Fig1:Potentials}
Illustration of three qualitatively different shapes for the potential in Eq.~(\ref{generalPotential}) for positive $b$: (a) $a>b$ with a lower minimum to the right of the origin and a real instanton; (b) $8b/9 < a < b$ with a higher minimum to the right of the origin and a complex instanton; (c) $a < 8b/9$ with origin being the only local minimum of the potential and complex instantons governing the nonperturbative behavior.}
\end{figure}

\begin{figure}
\begin{centering}
\includegraphics[width=5.6cm]{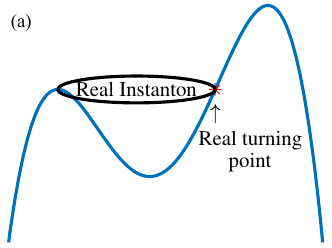} \hspace{1cm} \
\includegraphics[width=5.6cm]{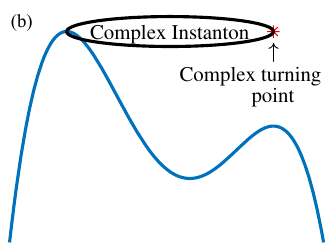}
\par\end{centering}
\caption{\label{Fig2:real-complex-instantons}
Illustration of
(a) real,
(b) complex instantons. }
\end{figure}

We will investigate a two-parameter perturbation of the harmonic oscillator:
\begin{equation} \label{generalPotential}
    V(x; a,b) = \frac12 x^2 - \sqrt{a} x^3 + \frac12 b x^4,
\end{equation}
which is the most general form of the genus-1 anharmonic oscillator and has the form of a tilted double-well for $b>0$. The minima of this potential for $b>0$ are at $x=0$ and at $x = \frac{\sqrt{9 a-8 b}+3 \sqrt{a}}{4 b}$, which lies to the right of the origin. For $a=b$, this potential reduces to the symmetric double-well potential with minima at $x = 0$ and $x = 1/\sqrt{a}$, the properties of which are well known. For $a>b$ [Fig.~\ref{Fig1:Potentials}(a)], the second minimum is below the minimum at the origin, and for an expansion around the local minimum at the origin, the nonperturbative properties of the system are governed by a real instanton [as in Fig.~\ref{Fig2:real-complex-instantons}(a)]. 

Complex instantons [Fig.~\ref{Fig2:real-complex-instantons}(b)] appear for $a<b$, when the origin is a global minimum of the potential. The potential is qualitatively different for $a > \frac{8}{9}b$ and $a < \frac{8}{9}b$. For $\frac{8}{9}b < a < b$, the potential has two minima, although the well to the right of the origin is tilted up [Fig.~\ref{Fig1:Potentials}(b)]; and for $a < \frac{8}{9}b$, the only local minimum of the potential is at the origin [Fig.~\ref{Fig1:Potentials}(c)]. Despite their qualitative differences, the contributions of complex instantons in both systems are of similar forms.

It should be noted that the systems depicted in Figs.~\ref{Fig1:Potentials}(a) and \ref{Fig1:Potentials}(b) can also be interpreted as expansions around two different minima of the same tilted double-well, because the potentials in Figs.~\ref{Fig1:Potentials}(a) and \ref{Fig1:Potentials}(b) can be converted to each other with a reflection and appropriate scaling and shifting of the potential. Therefore, the effects of real or complex instantons can be observed in the same system, depending on which minimum the perturbative expansion is performed around.

We will keep our discussion general for an arbitrary $a$ and $b$, which means that our results are applicable for both bounded ($b>0$) and unbounded ($b<0$) systems.

The equation for energy conservation defines a hypersurface in the complexified phase space $\mathbb{C}^2$. The genus of the curve is determined by the degree of the polynomial potential. For the family of potentials we study, in Eq.~(\ref{generalPotential}), we have a genus-1 Riemann surface,
\begin{equation} \label{RiemannSurfaceEq}
    p^2(x,\xi;a,b) = 2m(\xi - V(x; a,b)),
\end{equation}
which is a torus with two punctures (the poles at infinity on both Riemann sheets). Here, $\xi$ is the modulus of the surface and physically corresponds to the energy of a particle in the potential in Eq.~(\ref{generalPotential}). In the following discussion, we set $m = 1$ without loss of generality.

A period $S_{\mathcal{C}}(\xi;a,b)$ on this surface is defined as the integral of $p$ over a closed contour $\mathcal{C}$:
\begin{equation} \label{periodDef}
    S_{\mathcal{C}}(\xi;a,b) \equiv \oint_{\mathcal{C}} p(x,\xi;a,b) \dd{x}.
\end{equation}
 The Riemann surface described by Eq.~(\ref{RiemannSurfaceEq}) is a torus with two punctures. Hence, there are three independent periods on the surface, associated with the canonical 1-cycles on the torus and another period around either one of the punctures. Any other closed contour on the Riemann surface can be written as a linear combination of three independent contours. Therefore, three independent periods constitute a basis of the space of periods on the Riemann surface defined by Eq.~(\ref{RiemannSurfaceEq}), which is mathematically a module over the ring $\mathbb{Z}$. Notice that for the case of the symmetric double-well ($a=b$) and the quartic oscillator ($a=0$), the number of independent periods reduces to two because of the symmetries of the potential. For the case of the cubic oscillator ($b=0$), the two punctures merge, also resulting in a reduction of the number of independent periods. The manifestation of these special cases in our parametric P/NP relation is that the P/NP relations for those cases only include one derivative term with respect to the deformation parameter $\hbar$, as in the P/NP relation in Eq.~(\ref{Dunne-Unsal}). This reduction becomes more transparent in the parametric P/NP relation, which will be explained in the following.

\subsection{Classical periods and the Picard-Fuchs equation}

\par{It is known that the number of linearly independent periods defined on a 1-dimensional complex manifold is equal to the number of linearly independent 1-forms \cite{Gulden_2019}. Since the Riemann surface of the classical momentum defined in Eq.~(\ref{RiemannSurfaceEq}) is a 1d complex manifold with local coordinate $x$, this theorem is applicable in our case, and in fact, to the Riemann surfaces of all such defined classical momenta, given that $V(x)$ has a finite number of singularities in the complex plane.}

A significant corollary of this theorem is that the defining function of the Riemann surface, $p(x,\xi)$, is an analytic function on the same surface, and therefore defines a natural 1-form, written as $p(x,\xi) \dd{x}$ in the local coordinates. The integral of this 1-form over closed contours on the Riemann surface is what we actually defined as the periods in Eq.~(\ref{periodDef}). The derivatives of $p(x,\xi)$ with respect to both $x$ and $\xi$ are also analytic on the same Riemann surface and define 1-forms. If there are $K$ independent cycles on the Riemann surface, there are also $K$ independent 1-forms, and given $K+1$ different 1-forms on the surface, there must exist a linear combination of them which equals to an exact form (linear independence of forms is defined up to an exact form), written as $\dd{h(x,\xi)}$ in the local coordinates, $h(x,\xi)$ being an analytic function of $x$. If one considers the first $K$ derivatives with respect to $\xi$ of $p(x,\xi) \dd{x}$, then their linear dependence implies
\begin{equation} \label{Picard-Fuchs_oneforms}
    \sum_{i=0}^{K} f_i(\xi) \pdv[i]{\xi}  p(x,\xi) \dd{x} = \dd{h(x,\xi)}.
\end{equation}
This equation can be used to find the coefficients $f_i(\xi)$, which reduces to a simple linear algebra problem with an appropriate Ansatz for $h(x)$, which is straightforward to find for simple $V(x)$ (see \cite{Mironov:2009uv,Gulden_2019,Klemm_2019,Codesido:2017dns} for examples). If one integrates this expression over any closed contour $\mathcal{C}$, the integral of the exact form vanishes, and one obtains a $K$'th order linear homogeneous ordinary differential equation for the periods, called the Picard-Fuchs equation:
\begin{equation} \label{Picard-Fuchs}
    \sum_{i=0}^{K} f_i(\xi;a,b) \pdv[i]{\xi}  S_{\mathcal{C}}(\xi;a,b) = 0,
\end{equation}
where we introduced the dependence on $a$ and $b$ in the notation, but the result is applicable for all $V(x)$.
The coefficients $f_i(\xi;a,b)$ of this equation are independent of the period, and all periods on the surface satisfy this equation. For more details on the Picard-Fuchs equation and its usage, we refer the readers to \cite{Klemm_2019,Mironov:2009uv,Gulden_2019}

\begin{figure}
\begin{centering}
\includegraphics[width=10cm]{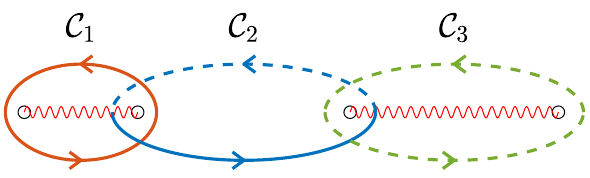}
\par\end{centering}
\caption{\label{Cycles}
Illustration of different contours on the complex $x$ plane, plotted for $a>b>0$ and $E >0$. Black circles are the zeros of $p(x)$, also called the turning points. Red wavy lines are the branch cuts of $p(x)$, and dashed lines indicate that the curve is in the second Riemann sheet. The integration of $p(x,\xi;a,b)$ over those contours gives $S_{\mathcal{C}}$. The first Riemann sheet is chosen such that $\text{Re}(p(x,\xi))$ is positive right below the origin for positive $\xi$, and as a result, $S_{\mathcal{C}_1}$ and $S_{\mathcal{C}_3}$ are positive and $S_{\mathcal{C}_2}$ is purely imaginary with positive imaginary part.}
\end{figure}

For the case of general genus-1 potentials, $K=3$, and the classical Picard-Fuchs equation in Eq.~(\ref{Picard-Fuchs}) is a third-order differential equation. Moreover, the coefficient of the zeroth derivative $f_0(\xi;a,b)$ vanishes; therefore, one of the solutions of the classical Picard-Fuchs equation is a constant. This is because the two \textit{perturbative} periods (i.e. the periods around the two minima of the potential in Eq.~(\ref{generalPotential}) which are nothing but the reduced actions for the classically allowed trajectories), which are the integrals over the contours $\mathcal{C}_1$ and $\mathcal{C}_3$ in Fig.~\ref{Cycles} and the analytic continuations thereof for other values of the parameters $a$ and $b$, differ only by a constant.

Besides the constant solution, we will define the two remaining classical periods as
\begin{equation}
    t(\xi;a,b) \equiv \frac{1}{2\pi} S_{\mathcal{C}_1}, \quad t_D(\xi;a,b) \equiv \frac{1}{i} S_{\mathcal{C}_2},
\end{equation}
where $\mathcal{C}_1$ is the contour around the turning points on the two sides of the minimum at the origin, corresponding to the perturbative period; and $\mathcal{C}_2$ is the contour enclosing the second and third turning points, corresponding to the nonperturbative period and encoding the tunneling information (Fig.~\ref{Cycles}).

These two periods can be calculated as power series solutions of the Picard-Fuchs equation in the modulus $\xi$. If we are looking for a solution around $\xi = 0$, only the classical perturbative period $t$ can be found by the substitution of the power series Ansatz $t = \sum_{i=0}^{\infty} a_i(a,b) \xi^i$, with $a_0=0$ and $a_1 = 1$.

The series expansion of the classical nonperturbative period contains logarithmic terms as well, and the coefficient of the logarithmic term can be determined by observing the behavior of the turning points as $\xi$ is rotated around $\xi = 0$ as $\xi \rightarrow \xi e^{2 \pi i}$. After this rotation, the two turning points around the origin swap places, which means that with the rotation of $\xi$, $S_{\mathcal{C}_2}$ deforms to $S_{\mathcal{C}_2} + S_{\mathcal{C}_1}$, which is only possible if $S_{\mathcal{C}_2}$ contains a term like $\frac{1}{2 \pi i} S_{\mathcal{C}_1}\log(\xi)$. This condition fixes the logarithmic term in $t_D$, and we can write an Ansatz for it as
\begin{equation} \label{t_D_Ansatz}
    t_D = t \log(\xi) + S(a,b) + b_1(a,b) t + \sum_{i=2}^{\infty} b_i(a,b) \xi^i,
\end{equation}
and the coefficients $b_i$ can be found by substituting this Ansatz into the Picard-Fuchs equation. 
The zeroth order term in $t_D$ is the instanton action $S$, defined as
\begin{equation}
    S = 2 \int_0^{x_3} \sqrt{2 V(x;a,b)} \dd{x},
\end{equation}
where $x_3 = \sqrt{a} - \sqrt{a-b}$. Notice the factor $2$ in front, which should be taken into account when comparing the results we present with the results in the literature, as the usual definition of the instanton action only includes the tunneling action. This factor in our definition corresponds to the inclusion of both the instanton and the anti-instanton action.

The evaluation of the instanton action with the potential in Eq.~(\ref{generalPotential}) gives
\begin{equation} \label{action}
    S = \frac{3 a-2 b}{3 b^2}+\frac{(a-b)}{2 b^2} \sqrt{\frac{a}{b}} \log \left(\frac{\sqrt{a}-\sqrt{b}}{\sqrt{a}+\sqrt{b}}\right).
\end{equation}

\subsection{Exact WKB method and the quantum periods}
So far, we have only considered the system classically. The transition to quantum mechanics is achieved through the time-independent Schrödinger equation, which can be put into the form of a Riccati equation by the substitution $\psi(x) = \exp(i \sigma(x,\xi;\hbar) / \hbar)$, which gives
\begin{equation}
\left(\sigma'(x,\xi;\hbar)\right)^2 - i \hbar  \sigma''(x,\xi;\hbar) = 2m\left(\xi - V(x)\right),
\end{equation}
which can be further reduced by the substitution $\rho(x,\xi;\hbar) = \sigma'(x,\xi;\hbar)$:
\begin{equation} \label{Riccati}
\left(\rho(x,\xi;\hbar)\right)^2 - i \hbar  \rho'(x,\xi;\hbar) = 2m\left(\xi - V(x)\right).
\end{equation}
With this substitution, the function $\rho(x,\xi;\hbar)$ can be regarded as a quantum generalization of the classical momentum $p(x,\xi)$, as the Riccati equation (\ref{Riccati}) reduces to the classical expression for the momentum in the limit $\hbar \rightarrow 0$ (as in Eq.~(\ref{RiemannSurfaceEq}), which can also be interpreted as the time-independent Hamilton-Jacobi equation), and in the general case, the second term on the left-hand side of Eq.~(\ref{Riccati}) introduces the quantum corrections, as it is the only term where $\hbar$ appears.

In the exact WKB approach, the Riccati equation (\ref{Riccati}) is solved by a series Ansatz for $\rho(x,\xi;\hbar)$ in $\hbar$:
\begin{equation}
    \rho(x,\xi;\hbar) = \sum_{k=0}^{\infty} \rho_k(x,\xi) \hbar^{k},
\end{equation}
which is substituted in Eq.~(\ref{Riccati}) and solved separately for each order in $\hbar$, which results in a recursion relation for the functions $\rho_k(x,\xi)$:
\begin{equation} \label{recursionWKB}
    \rho_k(x,\xi) = \frac{1}{2\rho_0(x,\xi)} \left(i \rho'_{k-1}(x,\xi) - \sum_{l=1}^{k-1}\rho_l(x,\xi)\rho_{k-l}(x,\xi) \right).
\end{equation}
where $\rho_0(x,\xi) = p(x,\xi)$, as can be seen from the zeroth order term in Eq.~(\ref{Riccati}). In the equation for the classical momentum, there are two solutions, one the negative of the other, corresponding to the two Riemann sheets of the square root function. Therefore, there are also two solutions for $\rho(x,\xi;\hbar)$, corresponding to $\rho_0 = p$ and $\rho_0 = -p$. From Eq.~(\ref{recursionWKB}), one can see that for $\rho_0 = -p$, the even orders in $\rho_k$ pick up a minus sign, whereas the odd orders in $\rho_k$ are the same for both $\rho_0 = -p$ and $\rho_0 = p$. Therefore, the two solutions of Eq.~(\ref{Riccati}) can be decomposed as
\begin{equation}
\begin{split}
    \rho^{\pm}(x,\xi;\hbar) &= \rho_{\text{odd}}(x,\xi;\hbar) \pm \rho_{\text{even}}(x,\xi;\hbar) \\
    &= \sum_{k=0}^{\infty} \rho_k(x,\xi) \hbar^{2k+1} \pm \sum_{k=0}^{\infty} \rho_k(x,\xi) \hbar^{2k},
\end{split}
\end{equation}
where both $\rho^+$ and $\rho^-$ solve the Riccati equation (\ref{Riccati}). Therefore, by substituting $\rho^+$ and $\rho^-$ in Eq.~(\ref{Riccati}) and taking the difference of the two equations, one can immediately find the relation
\begin{equation} \label{oddAndEvenQuantumPeriods}
\begin{split}
    \rho_{\text{odd}}(x,\xi;\hbar) &= \frac{i \hbar}{2} \frac{\rho'_{\text{even}}(x,\xi;\hbar)}{\rho_{\text{even}}(x,\xi;\hbar)} \\ 
    &= \frac{i \hbar}{2} \pdv{x} \log(\rho_{\text{even}}(x,\xi;\hbar)),
\end{split}
\end{equation}
which means that the knowledge of $\rho_{\text{even}}$ suffices to determine $\rho(x,\xi)$ completely.

Using the function $\rho(x,\xi)$, one can also write a quantization condition using the general properties of the wave functions of bound states. By definition, $\rho(x,\xi)$ can be written as
\begin{equation}
    \rho(x,\xi) = -i\hbar \frac{1}{\psi(x)} \dv{\psi(x)}{x}.
\end{equation}
From the Schrödinger equation, it is known that for the $N$'th excited state, $\psi(x)$ has $N$ zeros on the real axis, all of which lie in the classically accessible regions (in the regions where $V(x) < \xi$). Therefore, a closed contour enclosing all of the turning points of the classical momentum $p$ will also enclose the poles of $\rho(x,\xi)$, corresponding to the zeros of $\psi(x)$. The residue of each such pole will be $2\pi i$, and we can immediately write the quantization condition
\begin{equation}
    \oint_{\mathcal{C}}\rho(x,\xi;\hbar) \dd{x} = 2 \pi N \hbar.
\end{equation}
Due to Eq.~(\ref{oddAndEvenQuantumPeriods}), we can replace $\rho(x,\xi;\hbar)$ by $\rho_{\text{even}}(x,\xi;\hbar) + \frac{i \hbar}{2} \pdv{x} \log(\rho_{\text{even}}(x,\xi;\hbar))$, and integrate $\frac{i \hbar}{2} \pdv{x} \log(\rho_{\text{even}}(x,\xi;\hbar))$ over the contour $\mathcal{C}$ enclosing all the zeroes. Despite the integrand being a total derivative, this integration is not equal to zero, because although the contour we are integrating over is closed in the $\rho_{\text{even}}$-plane, it is not closed in the $\log(\rho_{\text{even}})$-plane, as the start and end points are in different Riemann sheets of the logarithm function, which introduces a difference of $2\pi i$. Therefore, we can write the quantization condition as
\begin{equation} \label{generalized_quantization_condition}
    \oint_{\mathcal{C}}\rho_{\text{even}}(x,\xi;\hbar) \dd{x} = 2 \pi \hbar \left(N+\frac{1}{2}\right),
\end{equation}
which is the famous generalized Bohr-Sommerfeld quantization condition.

The generalization of the classical momentum carries over to the classical periods, and the quantum periods are defined as
\begin{equation}
    \begin{split}
        \varsigma_{\mathcal{C}}(\xi;\hbar) &\equiv \oint_{\mathcal{C}} \rho_{\text{even}}(x,\xi;\hbar) \dd{x}, \\
        \varsigma_{\mathcal{C}}(\xi;\hbar) &= \sum_{k=0}^{\infty} \varsigma_{\mathcal{C},k}(x,\xi) \hbar^{2k},
    \end{split}
\end{equation}
where $\varsigma_{\mathcal{C},0} = S_{\mathcal{C}}$.

If, in Eq.~(\ref{generalized_quantization_condition}) one takes the contour $\mathcal{C}$ to be the perturbative contour ($\mathcal{C}_1$ in Fig.~\ref{Cycles}) instead of the contour enclosing the zeroes of the wave function, one obtains the perturbative quantization condition. Therefore, we can write the perturbative Bohr-Sommerfeld quantization condition for our potential in Eq.~(\ref{generalPotential}) as
\begin{equation} \label{Bohr-Sommerfeld}
    \varsigma_{\mathcal{C}_1}(\xi;a,b,\hbar) = 2\pi \hbar \left(N + \frac{1}{2} \right).
\end{equation}

Now that we have established the concept of the quantum periods, we need to be able to calculate them as well. Similarly to the calculation of the classical periods, we again resort to the theorem on the number of linearly independent 1-forms on the Riemann surface. From Eq.~(\ref{recursionWKB}), one can see that each $\rho_k(x,\xi)$ can be written in terms of $p(x,\xi)$ and its $x$-derivatives, and therefore each $\rho_k(x,\xi)$ is analytic on the Riemann surface of the classical momentum, and defines a 1-form. As explained in the previous section, we already know the number of linearly independent one-forms on the Riemann surface to be equal to $K$, and we can write $\rho_k(x,\xi) \dd{x}$ as
\begin{equation}
    \rho_{k}(x,\xi) \dd{x} = \sum_{i = 0}^{K-1} \mathcal{D}_{i,k}(\xi) \pdv[i+n_k]{\xi} p(x,\xi) \dd{x} + \dd{h_k(x,\xi)},
\end{equation}
This equation can again be solved for the coefficients $\mathcal{D}_{i,k}(\xi)$ with an appropriate Ansatz for $h_k(x,\xi)$. Here, $n_k$ can be chosen to be any positive integer but is usually chosen such that $\mathcal{D}_{i,k}(\xi)$ are as simple as possible.

Integrating this expression over a closed contour $\mathcal{C}$ and introducing the dependence on $a$ and $b$ for the case of our potential in Eq.~(\ref{generalPotential}), one obtains
\begin{equation}
    \varsigma_{\mathcal{C},k}(\xi;a,b) = \sum_{i = 0}^{K-1} \mathcal{D}_{i,k}(\xi;a,b) \pdv[i+n_k]{\xi} S_{\mathcal{C}}(\xi;a,b).
\end{equation}
The power of this formula is that once the coefficients $\mathcal{D}_{i,k}(\xi;a,b)$ are calculated, the quantum periods can immediately be calculated from the corresponding classical periods by taking derivatives, without having to solve new differential equations, and the coefficients $\mathcal{D}_{i,k}(\xi;a,b)$ are independent of the period.

With this result, the quantum periods can be written as a formal power series in $\hbar$:
\begin{equation}
    \varsigma_{\mathcal{C}}(\xi;a,b,\hbar) = S_{\mathcal{C}}(\xi;a,b) + \sum_{k=1}^{\infty} \sum_{i = 0}^{K-1} \hbar^{2k} \mathcal{D}_{i,k}(\xi;a,b) \pdv[i+n_k]{\xi} S_{\mathcal{C}}(\xi;a,b).
\end{equation}
In these equations, $\mathcal{D}_{i,k}$ are in general rational functions of $\xi$. If the periods are obtained as power series in $\xi$, the coefficients $\mathcal{D}_{i,k}$ also need to be expanded to a Taylor series, and the quantum periods are obtained as a double power series in $\xi$ and $\hbar$.

Using this generalization, we can also define the perturbative and nonperturbative quantum periods as
\begin{equation}
    \nu(\xi;a,b,\hbar) \equiv \frac{1}{2\pi} \varsigma_{\mathcal{C}_1}, \quad \nu_D(\xi;a,b,\hbar) \equiv \frac{1}{i} \varsigma_{\mathcal{C}_2},
\end{equation}
which respectively reduce to $t$ and $t_D$ as $\hbar \rightarrow 0$.

\subsection{Instanton function A and the parametric P/NP relation}

With the knowledge of the quantum periods $\nu$ and $\nu_D$, we can define the instanton function $\Tilde{A}$ as
\begin{equation} \label{A_def}
    \Tilde{A}(\xi;a,b,\hbar) \equiv \nu_D - \frac{\hbar}{2} \log{2\pi} + \hbar \log{\Gamma \left(\frac12 - \frac{\nu}{\hbar} \right)} - \left[ \log(-e \hbar) + b_1\right] \nu,
\end{equation}
where $b_1$ is the coefficient appearing in front of $t$ in the Ansatz for $t_D$ in Eq.~(\ref{t_D_Ansatz}). In this definition, the branch cuts of the logarithm function are chosen such that $\Tilde{A}$ is real, except for possibly the instanton action $S$ coming from $\nu_D$. With the proper choice of the Riemann sheets, the singularities appearing in $\nu_D$ and $\log{\Gamma \left(\frac12 - \frac{\nu}{\hbar} \right)}$ cancel each other, and the resulting $\Tilde{A}$ function does not have any singularities in either $\xi$ or $\hbar$. Moreover, because of the $-b_1 \nu$ term appearing in the definition, the coefficient of the $\xi^1 \hbar^0$ term in $\Tilde{A}$ is zero, independent of what $b_1$ is. Therefore, $\Tilde{A}$ is a function that completely encodes the nonperturbative information of the system but without the singularities appearing in $\nu_D$.

If the quantum periods $\nu$ and $\nu_D$ are known as power series in $\xi$ and $\hbar$, in order to calculate $\Tilde{A}$ as a power series as well, $\log{\Gamma \left(\frac12 - \frac{\nu}{\hbar} \right)}$ needs to be replaced by its asymptotic expansion for large $\nu$
\begin{equation} \label{loggamma_expansion}
    \hbar \log{\Gamma \left(\frac12 - \frac{\nu}{\hbar} \right)} \sim - \nu \log(-\frac{\nu}{e \hbar}) + \frac{\hbar}{2} \log{2\pi} - \sum_{k=2}^{\infty} \frac{B_k(1/2)}{k (k-1)} \frac{\hbar^k}{\nu^{k-1}},
\end{equation}
where $B_k(x)$ are the Bernoulli polynomials. Substituting (\ref{loggamma_expansion}) into (\ref{A_def}), we can rewrite $\Tilde{A}$ as
\begin{equation}
    \Tilde{A}(\xi;a,b,\hbar) = \nu_D - b_1 \nu - \nu \log{\nu} - \sum_{k=2}^{\infty} \frac{B_k(1/2)}{k (k-1)} \frac{\hbar^k}{\nu^{k-1}}.
\end{equation}
This expression can be used to calculate $\Tilde{A}$ as a power series; either as a power series in $\nu$ and $\hbar$, by inverting $\nu$ as $\xi(\nu;\hbar)$ and substituting for $\xi$ and writing $\nu_D$ as a power series in $\nu$ and $\hbar$; or directly as a power series in $\xi$ and $\hbar$, by substituting the series expansions of $\nu_D(\xi;a,b,\hbar)$ and $\nu(\xi;a,b,\hbar)$, and expanding the $\log{\nu}$ and $\nu^{1-k}$ terms into a series in $\xi$. Usually, first inverting $\nu$ and writing $\xi$ as a power series in $\nu$, and then substituting into $\nu_D$ and expanding around $\nu=0$, and then calculating $\Tilde{A}$ in terms of $\nu$ is computationally simpler.

We now define the perturbative function $B$ and the nonperturbative instanton function $A$ with the conventions in \cite{Zinn-Justin:2004vcw,Zinn-Justin:2004qzw,Jentschura:2010zza}. This is achieved simply by writing $\nu$, $\Tilde{A}$, and $\xi$ in units of $\hbar$, with the replacements $\nu \rightarrow \hbar B$, $\Tilde{A} \rightarrow \hbar A$, and $\xi \rightarrow \hbar E$. Therefore, $A$ and $B$ are defined as
\begin{equation}
    A(E;a,b,\hbar) \equiv \frac{1}{\hbar} \Tilde{A}(\hbar E; a, b, \hbar),
\end{equation}
and
\begin{equation} \label{Bdef}
    B(E; a, b, \hbar) \equiv \frac{1}{\hbar} \nu(\hbar E; a, b, \hbar).
\end{equation}
With these definitions, the correspondence with the results in the literature, where the perturbative series are written in terms of a coupling constant $g$, with $\hbar$ set to 1, is more transparent, and $\hbar$ in $A$ and $B$ exactly corresponds to the perturbation parameter $g$ in the literature. Moreover, these two functions are a power series only in the parameter $\hbar$; the series in $\xi$ is mixed into the series in $\hbar$ with the replacement $\xi \rightarrow \hbar E$, and as a result, the coefficient of each term in $\hbar$ is a polynomial in $E$. These functions exactly correspond to the functions appearing in the double-well quantization condition in Eq.~(\ref{DoubleWellQuantizationCondition}).

It is possible to directly obtain the Rayleigh-Schrödinger perturbation series from the $B$ function defined in Eq.~(\ref{Bdef}). In terms of $B$, the perturbative Bohr-Sommerfeld quantization condition in Eq.~(\ref{Bohr-Sommerfeld}) becomes $B = N + 1/2$, and if $B$ is inverted and $E$ is written in terms of $B$ as $E(B;a,b,\hbar)$, the substitution of $B \rightarrow N + 1/2$ in $E(B;a,b,\hbar)$ produces the perturbation series for the $N$'th energy level, with $N=0$ corresponding to the ground state.

With $E$ written in terms of $B$, it is also possible to write $A$ in terms of $B$ as $A(B;a,b,\hbar) = A(E(B;a,b,\hbar); a,b,\hbar)$. Using the parametrizations of $E$ and $A$ in terms of $B$, it was shown in \cite{Dunne:2013ada,Dunne:2014bca} that there is an explicit relation between $E(B;\hbar)$ and $A(B;\hbar)$ for various genus-1 systems with two independent periods. This relation is what we called the P/NP relation in Eq.~(\ref{Dunne-Unsal}). For generic anharmonic genus-1 potential in Eq.~(\ref{generalPotential}), such as the tilted double-well, the P/NP relation between $E(B;a,b,\hbar)$ and $A(B;a,b,\hbar)$ can be generalized as
\begin{equation} \label{generalPNP}
  \boxed{  \frac{\partial E}{\partial B}=-\hbar \left[  \left( \frac{15}{2}a - \frac32 b \right)B +\left( \frac{15}{2} a - \frac92 b \right) a \frac{\partial A}{\partial a} +  \left( \frac{9}{2} a - \frac32 b  \right) b \frac{\partial A}{\partial b} \right]}.
\end{equation}
Note that the P/NP relation in Eq.~(\ref{generalPNP}) produces the results known in the literature for the cubic potential \cite{Alvarez:2000, Gahramanov:2015yxk} at $a=g, b=0,\hbar=1$, and for the quartic potential \cite{Alvarez:2002, Gahramanov:2015yxk} at $a=0, b=g, \hbar = 1$. In fact, at closer inspection, one can notice that the coefficients appearing in this expression are nothing but the actions of the cubic and quartic oscillators, and one can write the parametric P/NP as
\begin{equation}
    \frac{\partial E}{\partial B}=-\hbar \left[  \left( \frac{1}{S_3}a + \frac{1}{S_4} b \right)B +\left( \frac{1}{S_3} a - \left(\frac{1}{S_3} + \frac{2}{S_4}  \right) b \right) a \frac{\partial A}{\partial a} +  \left( \left( \frac{1}{S_3} + \frac{2}{S_4} \right) a + \frac{1}{S_4} b  \right) b \frac{\partial A}{\partial b} \right],
\end{equation}
where $S_3 = 2/15$ is the action of the cubic oscillator, and $S_4 = - 2/3$ is the action of the quartic oscillator. Both of these values can be obtained from Eq.~(\ref{action}) after setting $a=1,b=0$ and $a = 0,b=1$, respectively.

\section{Large-order corrections}
In the previous section, we demonstrated the relationship between perturbative and nonperturbative generating functions for the unified genus-1 anharmonic potential. The relationship between these two functions, as shown in Eq.~(\ref{generalPNP}), is the manifestation of the link between the early terms of their respective expansions. Moreover, it is well known in the literature that corrections to the large-order growth of the perturbative series are governed by the early terms of the fluctuations around the instantons \cite{Bender:1969si, Bender:1973rz}. In the subsequent discussion, we elucidate the exact relationship between these two expansions, both for real and complex instantons.

To simplify our discussion, choose our potential of the form 
\begin{equation} \label{gammaPotential}
    V(x) = \frac12 x^2 - \gamma x^3 + \frac12 x^4
\end{equation}
where we set $a = \gamma^2$ and $b=1$.

The P/NP relation for this potential, which can be obtained by a change of variables in the general P/NP as $a \rightarrow \gamma^2 g$ and $b \rightarrow g$ (with $g$ taking on the role of $\hbar$ after setting $\hbar$ to $1$), reads:
\begin{equation} \label{gammaRealPNP}
\frac{\partial E}{\partial B}=-\frac32 g \left[ \left(5\gamma^2-1\right)B + \left(3\gamma^2-1\right)g\frac{\partial A}{\partial g} + \left(\gamma^2-1\right)\gamma \frac{\partial A}{\partial \gamma} \right].
\end{equation}
In this form, the connection with the P/NP relation of the double-well potential is more transparent, as the derivative term with respect to $\gamma$ vanishes after setting $\gamma \rightarrow 1$, and one obtains the P/NP relation in the literature \cite{Dunne:2013ada}. In this form, the P/NP equation is also more suitable for calculating $A$ from $B$, as after substituting an Ansatz for $A$ as a power series in $g$, one only needs to solve an ordinary differential equation in each order in $g$ (see App.~\ref{App:PNP-solution} for details).

\subsection{Real instantons}
We begin our discussion with real instantons. In the case of the potential in Eq.~(\ref{gammaPotential}), real instantons appear for $|\gamma|>1$. For simplicity, we set $\gamma=1.05$ in the subsequent discussion when discussing real instantons, but the results are similar for all values of the parameter $\gamma$ greater than $1$. As shown in Fig.~\ref{Fig2:real-complex-instantons}(a), the system then has a local minimum at the origin, and the perturbative expansion around it is nonalternating, grows factorially, and hence is divergent and non-Borel summable. We compute the perturbative energy levels using the BenderWu \textit{Mathematica} package of \cite{Sulejmanpasic:2016fwr}. For the ground state, the early terms of the energy read
\begin{equation}
    E_{0}(g) = \frac{1}{2} - \frac{3651}{3200}\,g - \frac{6692973}{1024000}\,g^2 - \frac{678789271989}{8192000000}\,g^3 - \cdots \, .
\end{equation}
On the other hand, the coefficients of the large-order terms in the perturbative expansion of the ground state energy level grow factorially in the form 
\begin{equation}\label{real_instanton:large_order_prediction_leading}
    a_n \sim -\frac{1}{\pi^{3/2}}\Gamma\left(n+\frac{1}{2}\right)(\gamma^2-1)^{-1/2}S_{\text{R}}^{-n-1/2} \, ,
\end{equation}
where $|\gamma| > 1$ and $S_{\text{R}}$ denotes the action of the real instanton, which can be computed using Eq.~(\ref{action}). The divergence of the perturbative expansion is intricately related to the nonperturbative contributions to the energy levels, namely the tunneling effects from the false vacuum, which can be accounted for by instanton effects. By taking those instanton contributions to the energy levels, one can then obtain the full energy spectrum, as can be written in a transseries form \cite{Zinn-Justin:2004qzw, Zinn-Justin:2004vcw, Jentschura:2010zza}.

Now we turn to the P/NP relation in Eq.~(\ref{gammaRealPNP}). The power of this formula is the ability to generate the nonperturbative contributions directly from the perturbative function $E(B,g)$. To see that, let us first express the energy in terms of the parameters $B$ and $g$
\begin{align}
    E(B, g) &= B- g\left(\frac{1083 }{320}B^2+\frac{1887}{6400}\right) - g^2 \left(\frac{12089821 }{512000}B^3+\frac{14682071 }{2048000}B\right) \nonumber \\
    & \quad- g^3 \left(\frac{522955121271 }{1638400000}B^4+\frac{705443318331 }{3276800000}B^2+\frac{1191419562159}{131072000000}\right) + \dots \, ,
\end{align}
which results in
\begin{equation}
    \frac{\partial E}{\partial B} = 1 - g\frac{1083 }{160} B - g^2\left(\frac{36269463 }{512000}B^2+\frac{14682071}{2048000}\right) - g^3\left(\frac{522955121271 }{409600000}B^3+\frac{705443318331 }{1638400000}B \right) - \, \dots \, .
\end{equation}
By solving Eq.~(\ref{gammaRealPNP}) (see App.~\ref{App:PNP-solution} for details) we obtain
\begin{align}
    A(B,g) &= \frac{1}{g} \left(\frac{20920-2583 \log (41)}{48000}\right) + g\left(\frac{3681821 }{131200}B^2+\frac{1666711}{1574400}\right) + g^2\left(\frac{919562432837 }{8606720000}B^3+\frac{2357376906257 }{34426880000}B\right)  \nonumber \\
    &\qquad + g^3 \left(\frac{19586141576910927067 }{6775209984000000}B^4+\frac{19199980026766993817 }{13550419968000000}B^2+\frac{153777594953280672709}{1626050396160000000}\right) + \dots \, . 
\end{align}
It is worth mentioning that the same result for the nonperturbative function $A(B,g)$ can alternatively be obtained by solving the Picard-Fuchs equation, as explained in Section \ref{Section:PFEquationAndPNP}. 

\begin{figure}[t]
    \centering
    \includegraphics[width=10cm]{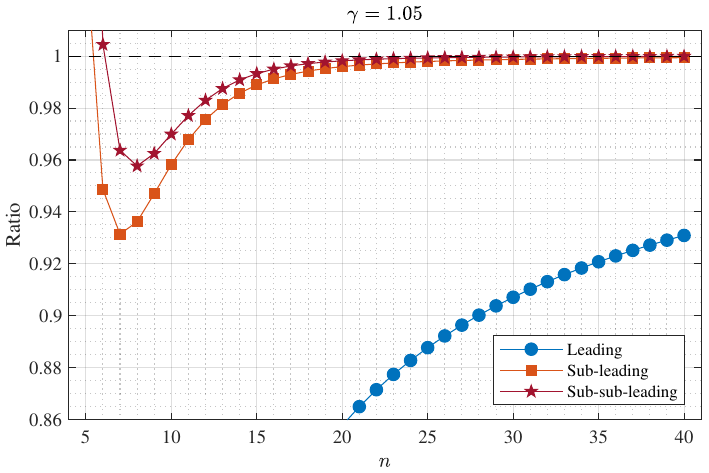}
    \caption{\label{Fig:real_instanton_comparison} In the presence of real instantons, ratios of the exact coefficients in the perturbative series with the leading order prediction (blue) and with the predictions with subleading (red) and sub-subleading (maroon) corrections included, up to order $n=40$.}
    \label{fig:real-instanton}
\end{figure}

With the nonperturbative function $A(B,g)$ one can easily compute the instanton fluctuation factor by \cite{Dunne:2013ada,Dunne:2014bca}
\begin{align}
    F(B, g) &= \text{exp}\left[-A(B,g) + \frac{S_{\text{R}}}{g}\right]\frac{\partial E(B,g)}{\partial B} \nonumber \\
    &= 1 + g\bigg(-\frac{3681821 }{131200}B^2-\frac{1083 }{160}B-\frac{1666711}{1574400}\bigg)  \nonumber \\
    &\quad \quad +g^2\bigg(\frac{13555805876041 }{34426880000}B^4+\frac{715276545793 }{8606720000}B^3-\frac{8496020591989 }{206561280000}B^2-\frac{2110687011147 }{34426880000}B \nonumber \\ 
    &\quad \qquad\qquad-\frac{32762082787919}{4957470720000}\bigg) + \dots \, .
\end{align}
The function $F(B,g)$ is the fluctuation around a single instanton and introduces corrections to the large-order growth in Eq.~(\ref{real_instanton:large_order_prediction_leading}) of perturbative series via 
\begin{equation}\label{real_instanton:large_order_prediction}
    a_n = -\frac{1}{\pi^{3/2}}\Gamma\left(n+\frac{1}{2}\right)(\gamma^2-1)^{-1/2}S_{\text{R}}^{-n-1/2} \bigg(1+\frac{b_1\,S_{\text{R}}}{n-\frac12}+\frac{b_2\,S_{\text{R}}^2}{\left(n-\frac12\right) \left(n-\frac32\right)}+\dots \bigg) \, ,
\end{equation}
where the coefficients $b_n$ represent the coefficients of $g^n$ in $F(B,g)$ when $B=N + 1/2$. For the ground state ($N = 0$), the first three of the coefficients read
\begin{align}
    b_1 &= -\frac{9020267}{787200} \, , \nonumber \\
    b_2 &= -\frac{15551376740243}{1239367680000} \, , \nonumber \\
    b_3 &= -\frac{6676028154247025473771}{14634453565440000000} \, . 
\end{align}

We validate our results by comparing the ratios of the exact coefficients obtained from the Bender-Wu \textit{Mathematica} package \cite{Sulejmanpasic:2016fwr} in the ground state perturbative series with those obtained from the leading order prediction Eq.~(\ref{real_instanton:large_order_prediction_leading}) as well as the leading order plus fluctuation terms as specified in Eq.~(\ref{real_instanton:large_order_prediction}). Figure~\ref{Fig:real_instanton_comparison} illustrates that the significant improvement of the numerical results can be achieved by including the fluctuation terms, starting from the very early terms of the series. We compare leading order results with subleading and sub-subleading results. Already at $n=20$, our results show a ratio of 0.998 at sub-subleading order while excluding the fluctuation terms results in a ratio of 0.858. This result further confirms our main finding, Eq.~(\ref{generalPNP}), and demonstrates the power of resurgence in generating the instanton fluctuation terms exclusively from perturbative information, thereby generating the large-order correction terms of the perturbative series.

\subsection{Complex instantons}

Next, we discuss the case where $|\gamma| < 1$, i.e., complex instantons are present in the system. The role of complex instantons has been extensively studied in numerous studies \cite{Cherman:2014ofa, Behtash:2015loa, Behtash:2015zha, Fujimori:2016ljw, Fujimori:2017oab, Fujimori:2017osz, Fujimori:2018kqp, Fujimori:2022lng, Dunne:2016jsr, Kozcaz:2016wvy, Dorigoni:2017smz, Dorigoni:2019kux}. This is a special case because, as opposed to the real instantons we discussed in the previous section, those solutions have no physical tunneling effect on the system. In such systems, the perturbative expansion is asymptotic and Borel summable, as the singularities of the Borel transform are not on the real axis \cite{Brezin:1976wa, Marino:2015yie}. Despite the absence of an actual tunneling effect from the global minimum, it is well known in the literature that complex instantons continue to govern the large-order behavior of the perturbative saddle \cite{Brezin:1976wa, Marino:2015yie}. In this section, we expand on those studies by generating the fluctuation terms to the large-order behavior of perturbative series using the P/NP relation in Eq.~(\ref{gammaRealPNP}). We will demonstrate that in systems with complex instantons, the fluctuation terms are extremely crucial, in fact, fundamental, in predicting the large-order behavior of perturbative series. 

For the potential in Eq.~(\ref{gammaPotential}), the minimum at the origin becomes the global minimum of the potential for $\abs{\gamma} < 1$, and although there is no physical tunneling  effect for particles at the origin and energies close to $0$, the complex instantons still contribute to the large-order growth of the perturbative series. To investigate this effect, we set $\gamma = 0.95$ for the following discussion. The early terms of the ground state energy, then, read
\begin{equation}
    E_{0}(g) = \frac{1}{2} - \frac{2771}{3200}\,g - \frac{2914893}{1024000}\,g^2 - \frac{157753309829}{8192000000}\,g^3 - \cdots \, .
\end{equation}
On the other hand, the terms in the perturbative ground state energy grow factorially in the form 
\begin{equation}\label{complex_instanton:large_order_prediction_leading}
    a_n \sim \frac{2}{\pi^{3/2}}\Gamma \left(n+\frac{1}{2} \right)(1-\gamma^2)^{-1/2} \, \text{Im} \big(S_{\text{C}}^{-n-1/2} \big) \, ,
\end{equation}
where $S_{\text{C}}$ is the action of the complex instanton, which has a complex value and can be readily calculated using Eq.~(\ref{action}). The imaginary part of Eq.~(\ref{complex_instanton:large_order_prediction_leading}) is a consequence of the presence of two individual instanton contributions, which manifest as complex conjugate pairs, and dictate the large-order growth of the perturbative expansion.

The energy expression is found to be
\begin{align}
    E(B,g) &= B - g \left(\frac{843 }{320}B^2+\frac{1327}{6400}\right) - g^2 \left(\frac{5923261 }{512000}B^3+\frac{5736311 }{2048000}B\right) \nonumber \\
    & \quad- g^3 \left(\frac{159756605031 }{1638400000}B^4+\frac{159553091691 }{3276800000}B^2+\frac{129739015199}{131072000000}\right) + \dots\,,
\end{align}
which leads to
\begin{equation}\label{complex_dE_dB}
    \frac{\partial E}{\partial B} = 1 -g\frac{843 }{160}B - g^2\left(\frac{17769783 }{512000}B^2+\frac{5736311}{2048000}\right)  - g^3\left(\frac{159756605031 }{409600000}B^3+\frac{159553091691 }{1638400000}B\right)-\dots \, .
\end{equation}
Inserting the parametric form of Eq.~(\ref{complex_dE_dB}) in Eq.~(\ref{gammaRealPNP}) and solving it for $A(B,g)$ then yields (see App.~\ref{App:PNP-solution} for details)
\begin{align}
    A(B,g) &= \frac{1}{g} \frac{11320 +2223 \left(\log (39) \pm i \pi\right)}{48000} + g\bigg(\frac{564739 }{124800}B^2+\frac{3362249}{1497600}\bigg)  + g^2\bigg(\frac{339564114191 }{23362560000}B^3+\frac{4180279643251 }{93450240000}B\bigg) \nonumber \\
    & \qquad +g^3\bigg(\frac{188164707059018693 }{5831294976000000}B^4+\frac{10650892452475311943 }{11662589952000000}B^2-\frac{14589991555431385189}{1399510794240000000}\bigg) + \dots \, .
\end{align}

\begin{figure}[t]
\begin{centering}
\includegraphics[width=\textwidth]{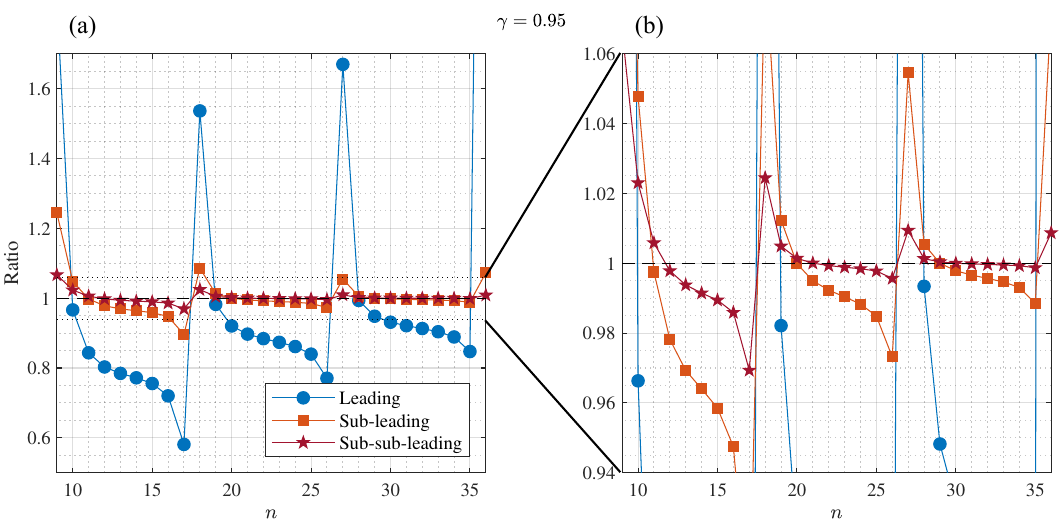}
\par\end{centering}
\caption{\label{Fig:complex_instanton_comparison} 
In the presence of complex instantons, comparison of the ratios of the exact coefficients in the perturbative series with the leading order prediction (blue) and with the predictions with subleading (red) and sub-subleading (maroon) corrections included, up to order $n = 35$.}
\end{figure}

With the nonperturbative function $A(B,g)$ one can easily compute the instanton fluctuation factor by 
\begin{align}
    F(B, g) &= \text{exp}\left[-A(B,g) + \frac{S_\text{C}}{g}\right] \frac{\partial E(B,g)}{\partial B} \nonumber \\
    & = 1 + g\bigg(-\frac{564739 }{124800}B^2-\frac{843 }{160}B-\frac{3362249}{1497600}\bigg) \nonumber \\
    &\quad\quad+g^2\bigg(\frac{318930138121 }{31150080000}B^4+\frac{217443608899 }{23362560000}B^3-\frac{4587888448309 }{186900480000}B^2-\frac{3074873039521 }{93450240000} B\nonumber \\
    &\qquad\qquad\quad-\frac{1259179466639}{4485611520000}\bigg) + \dots \, ,
\end{align}
The function $F(B,g)$ gives the fluctuation around a complex instanton and is related to the large-order growth of the perturbative series via
\begin{equation}\label{complex_instanton:large_order_prediction}
    a_n =\frac{2}{\pi^{3/2}}\Gamma \left(n+\frac{1}{2} \right)(1-\gamma^2)^{-1/2} \, \text{Im} \Bigg[S_{\text{C}}^{-n-1/2} \bigg(1+\frac{b_1\,S_{\text{C}}}{n-\frac12}+\frac{b_2\,S_{\text{C}}^2}{\left(n-\frac12\right) \left(n-\frac32\right)}+\dots \bigg) \Bigg] \, ,
\end{equation}
with coefficients $b_n$ representing the coefficients of $g^n$ in $F(B,g)$ when $B=N + 1/2$. In the ground state ($N =0$), the first three of them are
\begin{align}
    b_1 &= -\frac{4500853}{748800} \, , \nonumber \\
    b_2 &= -\frac{23623611312083}{1121402880000} \, , \nonumber \\
    b_3 &= -\frac{1948157258453830813109}{12595597148160000000} \, . 
\end{align}

To validate our results, we compare the ratios of the exact coefficients obtained from the Bender-Wu \textit{Mathematica} package \cite{Sulejmanpasic:2016fwr} in the ground state perturbative series with those derived from the leading order prediction Eq.~(\ref{complex_instanton:large_order_prediction_leading}), as well as the leading order plus fluctuation terms as described in Eq.~(\ref{complex_instanton:large_order_prediction}). Figure~\ref{Fig:complex_instanton_comparison}(a) illustrates the ratio between the exact coefficients and the leading order prediction as specified in Eq.~(\ref{complex_instanton:large_order_prediction_leading}). In the same plot, we also compare the ratios of including subleading and sub-subleading corrections. Omitting the fluctuation terms clearly illustrates the fundamental role of these coefficients in predicting the large-order prediction, as the outliers evident in the figure show. Incorporating these fluctuation terms eliminates outliers and the ratio of the exact values and predictions becomes centered around the value of 1, as seen in Fig.~\ref{Fig:complex_instanton_comparison}(b).

\section{Conclusions}
\par{The P/NP relation distinguishes itself from the rest of the other known resurgence methods by being constructive as it generates nonperturbative quantum fluctuations around an instanton from merely the knowledge of the perturbative expansion to the same order. Combined with the general quantization condition, it provides a very powerful approach to construct the exact energy levels of one dimensional quantum mechanical systems in the form of transseries. The P/NP relation has been tested only for a handful of specific potentials before. To the best of our knowledge, our parametric form of this relationship in Eq.~(\ref{generalPNP}) is the first instance where it was generalized to a continuous family of potentials expressing the connection between the perturbative and nonperturbative quantities in terms of deformation parameters. In the proper limits, our expression reproduces the known P/NP relation; hence, it includes the potentials studied earlier, like cubic, quartic, and symmetric double-well. }
\par{Unfortunately, the P/NP relation is restricted to the potentials that give rise to genus-1 hyperelliptic curves. The generalization for higher genera, i.e. generating independent nonperturbative fluctuation expansions solely from perturbative expansions, has been an interesting problem, which has been an open question for almost a decade now. We hope that our approach to study the family of potentials of a given genus could offer a resolution for this long standing problem.}
\par{Our deformation breaks the parity symmetry of the problem and allows us to study the perturbative expansion around the global or local minimum of the tilted potential. Depending on the minimum in question, either the real or the complex instantons govern the large-order behavior around that minimum. For both cases, we compared the perturbative expansion coefficients for the ground state, which we obtained from the Bender-Wu \textit{Mathematica} package with the leading order predictions and the fluctuation terms. We showed how including the fluctuation terms has improved the ratio between these distinct approaches. However, we emphasize that the improvement due to the fluctuation terms is more remarkable for the complex instantons. Due to the nontrivial phase of the complex instanton action, the leading expansion exhibits significant deviations from the actual one. After we include only a few correction terms, the agreement becomes considerable. Our results constitute the first example of the inclusion of fluctuation terms for the complex instantons.   }

\vspace{1cm}

{\bf Acknowledgments.}
We express our gratitude to Mithat \"Unsal for engaging in numerous discussions with us. We are also grateful to Ilmar Gahramanov, Charlotte Kristjansen, and Mustafa T\"ure for various discussions. C.K. would like to thank the warm hospitality of Niels Bohr Institute where this work was finished. The work of C.K. was supported by the Scientific and Technological Research Council of Turkey (TUBITAK) under the Grant Numbers 120F184 and 220N106. CK thanks TUBITAK for their support. This article is based upon work from COST Action 21109 CaLISTA and COST Action 22113 THEORY-CHALLENGES, supported by COST (European Cooperation in Science and Technology).

\appendix

\section{Calculation of A from B using the P/NP relation}\label{App:PNP-solution}

For the potential in Eq.~(\ref{gammaPotential}), if the perturbative expansion of $E$ in terms of $g$, $B$, and $\gamma$ is known, it is straightforward to calculate $A$ using the P/NP relation in Eq.~(\ref{gammaRealPNP}). For this potential, $E(B;g,\gamma)$ was calculated from the Picard-Fuchs equation as described in Section \ref{Section:PFEquationAndPNP}, and reads
\begin{align}
        E(B;g,\gamma) &= B + g \bigg[ \left(-\frac{15 \gamma ^2}{4}+\frac{3}{4}\right)B^2-\left(\frac{7 \gamma ^2}{16}+\frac{3}{16}\right)\bigg] \nonumber \\
        & + g^2 \bigg[ \left(-\frac{705 \gamma ^4}{16}+\frac{225 \gamma ^2}{8}-\frac{17}{16}\right)B^3+ \left(-\frac{1155 \gamma ^4}{64}+\frac{459 \gamma ^2}{32}-\frac{67}{64}\right)B\bigg]
        \nonumber \\
        & + g^3 \bigg[ \left(-\frac{115755 \gamma ^6}{128}+\frac{116325 \gamma ^4}{128}-\frac{24945 \gamma ^2}{128}+\frac{375}{128}\right) B^4+  \left(-\frac{209055 \gamma ^6}{256}+\frac{239985 \gamma ^4}{256}-\frac{62013 \gamma ^2}{256}+\frac{1707}{256}\right) B^2\nonumber \\
        & \qquad\quad -\left(\frac{101479 \gamma ^6}{2048}+\frac{131817 \gamma ^4}{2048}-\frac{40261 \gamma ^2}{2048}+\frac{1539}{2048}\right)\bigg]+\dots\,.
\end{align}
Note that this expression is nothing but the perturbative series for the energy levels after setting $B = N + \frac{1}{2}$. Therefore, it can also be easily calculated from standard perturbation theory, such as using the Bender-Wu \textit{Mathematica} package \cite{Sulejmanpasic:2016fwr}. 

To find $A(B;g,\gamma)$ as a power series, we write an Ansatz as
\begin{equation} \label{A_Ansatz}
    A(B;g,\gamma) = \frac{S(\gamma)}{g} + \sum_{i=1}^{\infty} f_{i}(B;\gamma) g^i\,,
\end{equation}
and insert this into the P/NP relation in Eq.~(\ref{gammaRealPNP}). To order $g^0$, we find a first-order ordinary differential equation for $S(\gamma)$:
\begin{equation}
    \frac{3}{2} \gamma  \left(\gamma ^2-1\right) \dv{S(\gamma)}{\gamma} -\frac{3}{2}\left(3 \gamma ^2-1\right) S(\gamma) + 1 = 0.
\end{equation}
The general solution of this equation is
\begin{equation}
    S(\gamma) = -\frac{2}{3} +\gamma^2+\frac{1}{2} \gamma \left(\gamma^2 -1\right) \log (\frac{\gamma-1}{\gamma+1}) + c_0 \gamma \left(\gamma^2 -1 \right),
\end{equation}
where $c_0$ is a constant that needs to be fixed. This can be done by calculating the instanton action for a specific $\gamma$ value, such as for $\gamma = 2$. Through such a calculation, it can be seen that $c_0 = 0$, which is consistent with the instanton action in Eq.~(\ref{action}), after setting $a = \gamma^2$ and $b = 1$. One can also see that $c_0=0$ from the condition that $S(\gamma)$ stays invariant under the reflection $\gamma \rightarrow -\gamma$, which is equivalent to the reflection $x \rightarrow -x$ in the potential in Eq.~(\ref{gammaPotential}).

To find the higher order terms of $A$, the differential equations obtained for the coefficients $f_i$ in the Ansatz for $A$ in Eq.~(\ref{A_Ansatz}) need to be solved. To order $g^{n+1}$ in the P/NP relation, we find a differential equation for $f_n(B;\gamma)$:
\begin{equation} \label{ADiffEq}
    \left(\gamma^2 -1 \right) \gamma \pdv{f_n(B;\gamma)}{\gamma} + n \left(3 \gamma^2 - 1\right) f_n(B;\gamma) = -\frac{2}{3} h_n(B;\gamma),
\end{equation}
where $h_n(B;\gamma)$ is the coefficient of the $g^{n+1}$ term in $\pdv*{\xi}{B}$.

The homogeneous solution of the differential equation above is
\begin{equation}
    f_n^{(\text{hom.})}(\gamma) = \frac{1}{\gamma^n (\gamma^2-1)^n},
\end{equation}
and the general solution becomes 
\begin{equation} \label{AdiffEqSol}
    f_n(B;\gamma) = f_n^{(\text{inhom.})}(B;\gamma) + c_n(B)f_n^{(\text{hom.})}(\gamma)
\end{equation}
Here, $f_n^{(\text{inhom.})}(B;\gamma)$ is the inhomogeneous solution of the differential equation (\ref{ADiffEq}), which can easily be found by the substitution $f_n^{(\text{inhom.})}(B;\gamma) \rightarrow \gamma^{-n} (\gamma^2-1)^{-n} g_n(B;\gamma)$, which transforms the differential equation (\ref{ADiffEq}) into
\begin{equation}
    \pdv{g_n(B;\gamma)}{\gamma} = -\frac{2}{3} \gamma ^{n-1} \left(\gamma ^2-1\right)^{n-1} h_n(B;\gamma),
\end{equation}
which can be integrated to find $f_n^{(\text{inhom.})}(B;\gamma)$, as the right-hand side is a polynomial in $\gamma$, starting from the $\gamma^{n-1}$ term.

The only remaining step is to fix the coefficients $c_n(B)$ of the homogeneous terms in Eq.~(\ref{AdiffEqSol}). From the condition that $A(B;g,\gamma)$ be regular for $\gamma = 0$, they can be easily seen to be equal to $0$ because the only terms with singularities of the form $\gamma^{-n}$ come from the homogeneous solutions.

After these steps, $A(B;g,\gamma)$ is found to be
\begin{align}
    A(B;g,\gamma) &= \frac{1}{g} \bigg[-\frac{2}{3} + \gamma ^2+\frac{1}{2} \left(\gamma ^2-1\right) \gamma  \log (\frac{\gamma -1}{\gamma+1})\bigg] \nonumber \\
    &+ \frac{g}{\gamma^2-1} \bigg[  \left(\frac{141 \gamma ^4}{8}-\frac{75 \gamma ^2}{4}+\frac{17}{8}\right)B^2+\left(\frac{77 \gamma ^4}{32}-\frac{51 \gamma ^2}{16}+\frac{67}{96}\right)\bigg] \nonumber \\ 
    &+ \frac{g^2}{(\gamma^2-1)^2} \bigg[  \left(\frac{7717 \gamma ^8}{32}-\frac{4835 \gamma ^6}{8}+\frac{23545 \gamma ^4}{48}-\frac{1055 \gamma ^2}{8}+\frac{125}{32}\right)B^3 \nonumber \\
    & \qquad\qquad\qquad\quad +  \left(\frac{13937 \gamma ^8}{128}-\frac{9355 \gamma ^6}{32}+\frac{50333 \gamma ^4}{192}-\frac{2655 \gamma ^2}{32}+\frac{569}{128}\right)B\bigg] + \dots \, .
\end{align}
The same result can also be obtained by solving the Picard-Fuchs equation and calculating the nonperturbative quantum period and using Eq.~(\ref{A_def}), as described in Section \ref{Section:PFEquationAndPNP}

These results can easily be extended to the case of the more general potential in Eq.~(\ref{generalPotential}) by the substitutions $g \rightarrow \hbar b$ and $\gamma \rightarrow \sqrt{a/b}$, and the resulting functions become power series in $\hbar$.

\bibliography{refs}

\begin{thebibliography}{45}%
\makeatletter
\providecommand \@ifxundefined [1]{%
 \@ifx{#1\undefined}
}%
\providecommand \@ifnum [1]{%
 \ifnum #1\expandafter \@firstoftwo
 \else \expandafter \@secondoftwo
 \fi
}%
\providecommand \@ifx [1]{%
 \ifx #1\expandafter \@firstoftwo
 \else \expandafter \@secondoftwo
 \fi
}%
\providecommand \natexlab [1]{#1}%
\providecommand \enquote  [1]{``#1''}%
\providecommand \bibnamefont  [1]{#1}%
\providecommand \bibfnamefont [1]{#1}%
\providecommand \citenamefont [1]{#1}%
\providecommand \href@noop [0]{\@secondoftwo}%
\providecommand \href [0]{\begingroup \@sanitize@url \@href}%
\providecommand \@href[1]{\@@startlink{#1}\@@href}%
\providecommand \@@href[1]{\endgroup#1\@@endlink}%
\providecommand \@sanitize@url [0]{\catcode `\\12\catcode `\$12\catcode `\&12\catcode `\#12\catcode `\^12\catcode `\_12\catcode `\%12\relax}%
\providecommand \@@startlink[1]{}%
\providecommand \@@endlink[0]{}%
\providecommand \url  [0]{\begingroup\@sanitize@url \@url }%
\providecommand \@url [1]{\endgroup\@href {#1}{\urlprefix }}%
\providecommand \urlprefix  [0]{URL }%
\providecommand \Eprint [0]{\href }%
\providecommand \doibase [0]{https://doi.org/}%
\providecommand \selectlanguage [0]{\@gobble}%
\providecommand \bibinfo  [0]{\@secondoftwo}%
\providecommand \bibfield  [0]{\@secondoftwo}%
\providecommand \translation [1]{[#1]}%
\providecommand \BibitemOpen [0]{}%
\providecommand \bibitemStop [0]{}%
\providecommand \bibitemNoStop [0]{.\EOS\space}%
\providecommand \EOS [0]{\spacefactor3000\relax}%
\providecommand \BibitemShut  [1]{\csname bibitem#1\endcsname}%
\let\auto@bib@innerbib\@empty
\bibitem [{\citenamefont {Zinn-Justin}\ and\ \citenamefont {Jentschura}(2004{\natexlab{a}})}]{Zinn-Justin:2004vcw}%
  \BibitemOpen
  \bibfield  {author} {\bibinfo {author} {\bibfnamefont {J.}~\bibnamefont {Zinn-Justin}}\ and\ \bibinfo {author} {\bibfnamefont {U.~D.}\ \bibnamefont {Jentschura}},\ }\bibfield  {title} {\bibinfo {title} {{Multi-instantons and exact results I: Conjectures, WKB expansions, and instanton interactions}},\ }\href {https://doi.org/10.1016/j.aop.2004.04.004} {\bibfield  {journal} {\bibinfo  {journal} {Ann. Phys. (N.Y.)}\ }\textbf {\bibinfo {volume} {313}},\ \bibinfo {pages} {197} (\bibinfo {year} {2004}{\natexlab{a}})},\ \Eprint {https://arxiv.org/abs/quant-ph/0501136} {arXiv:quant-ph/0501136} \BibitemShut {NoStop}%
\bibitem [{\citenamefont {Zinn-Justin}\ and\ \citenamefont {Jentschura}(2004{\natexlab{b}})}]{Zinn-Justin:2004qzw}%
  \BibitemOpen
  \bibfield  {author} {\bibinfo {author} {\bibfnamefont {J.}~\bibnamefont {Zinn-Justin}}\ and\ \bibinfo {author} {\bibfnamefont {U.~D.}\ \bibnamefont {Jentschura}},\ }\bibfield  {title} {\bibinfo {title} {{Multi-instantons and exact results II: Specific cases, higher-order effects, and numerical calculations}},\ }\href {https://doi.org/10.1016/j.aop.2004.04.003} {\bibfield  {journal} {\bibinfo  {journal} {Ann. Phys. (N.Y.)}\ }\textbf {\bibinfo {volume} {313}},\ \bibinfo {pages} {269} (\bibinfo {year} {2004}{\natexlab{b}})},\ \Eprint {https://arxiv.org/abs/quant-ph/0501137} {arXiv:quant-ph/0501137} \BibitemShut {NoStop}%
\bibitem [{\citenamefont {Jentschura}\ \emph {et~al.}(2010)\citenamefont {Jentschura}, \citenamefont {Surzhykov},\ and\ \citenamefont {Zinn-Justin}}]{Jentschura:2010zza}%
  \BibitemOpen
  \bibfield  {author} {\bibinfo {author} {\bibfnamefont {U.~D.}\ \bibnamefont {Jentschura}}, \bibinfo {author} {\bibfnamefont {A.}~\bibnamefont {Surzhykov}},\ and\ \bibinfo {author} {\bibfnamefont {J.}~\bibnamefont {Zinn-Justin}},\ }\bibfield  {title} {\bibinfo {title} {{Multi-instantons and exact results. III: Unification of even and odd anharmonic oscillators}},\ }\href {https://doi.org/10.1016/j.aop.2010.01.002} {\bibfield  {journal} {\bibinfo  {journal} {Ann. Phys. (N.Y.)}\ }\textbf {\bibinfo {volume} {325}},\ \bibinfo {pages} {1135} (\bibinfo {year} {2010})},\ \Eprint {https://arxiv.org/abs/1001.3910} {arXiv:1001.3910 [math-ph]} \BibitemShut {NoStop}%
\bibitem [{\citenamefont {\'Ecalle}(1981)}]{Ecalle:1981}%
  \BibitemOpen
  \bibfield  {author} {\bibinfo {author} {\bibfnamefont {J.}~\bibnamefont {\'Ecalle}},\ }\href@noop {} {\emph {\bibinfo {title} {{Les Fonctions R\'esurgentes}}}},\ Vol.\ \bibinfo {volume} {I-III}\ (\bibinfo  {publisher} {Publications Math\'ematiques d'Orsay},\ \bibinfo {address} {Orsay, France},\ \bibinfo {year} {1981})\BibitemShut {NoStop}%
\bibitem [{\citenamefont {Sueishi}\ \emph {et~al.}(2020)\citenamefont {Sueishi}, \citenamefont {Kamata}, \citenamefont {Misumi},\ and\ \citenamefont {\"Unsal}}]{Sueishi:2020rug}%
  \BibitemOpen
  \bibfield  {author} {\bibinfo {author} {\bibfnamefont {N.}~\bibnamefont {Sueishi}}, \bibinfo {author} {\bibfnamefont {S.}~\bibnamefont {Kamata}}, \bibinfo {author} {\bibfnamefont {T.}~\bibnamefont {Misumi}},\ and\ \bibinfo {author} {\bibfnamefont {M.}~\bibnamefont {\"Unsal}},\ }\bibfield  {title} {\bibinfo {title} {{On exact-WKB analysis, resurgent structure, and quantization conditions}},\ }\href {https://doi.org/10.1007/JHEP12(2020)114} {\bibfield  {journal} {\bibinfo  {journal} {J. High Energy Phys.}\ }\textbf {\bibinfo {volume} {2020}}\bibfield  {number} {\bibinfo  {number} { (12)},\ \bibinfo {pages} {114}},\ }\Eprint {https://arxiv.org/abs/2008.00379} {arXiv:2008.00379 [hep-th]} \BibitemShut {NoStop}%
\bibitem [{\citenamefont {Sueishi}\ \emph {et~al.}(2021)\citenamefont {Sueishi}, \citenamefont {Kamata}, \citenamefont {Misumi},\ and\ \citenamefont {\"Unsal}}]{Sueishi:2021xti}%
  \BibitemOpen
  \bibfield  {author} {\bibinfo {author} {\bibfnamefont {N.}~\bibnamefont {Sueishi}}, \bibinfo {author} {\bibfnamefont {S.}~\bibnamefont {Kamata}}, \bibinfo {author} {\bibfnamefont {T.}~\bibnamefont {Misumi}},\ and\ \bibinfo {author} {\bibfnamefont {M.}~\bibnamefont {\"Unsal}},\ }\bibfield  {title} {\bibinfo {title} {{Exact-WKB, complete resurgent structure, and mixed anomaly in quantum mechanics on S$^{1}$}},\ }\href {https://doi.org/10.1007/JHEP07(2021)096} {\bibfield  {journal} {\bibinfo  {journal} {J. High Energy Phys.}\ }\textbf {\bibinfo {volume} {2021}}\bibfield  {number} {\bibinfo  {number} { (07)},\ \bibinfo {pages} {096}},\ }\Eprint {https://arxiv.org/abs/2103.06586} {arXiv:2103.06586 [quant-ph]} \BibitemShut {NoStop}%
\bibitem [{\citenamefont {Kamata}\ \emph {et~al.}(2023)\citenamefont {Kamata}, \citenamefont {Misumi}, \citenamefont {Sueishi},\ and\ \citenamefont {\"Unsal}}]{Kamata:2021jrs}%
  \BibitemOpen
  \bibfield  {author} {\bibinfo {author} {\bibfnamefont {S.}~\bibnamefont {Kamata}}, \bibinfo {author} {\bibfnamefont {T.}~\bibnamefont {Misumi}}, \bibinfo {author} {\bibfnamefont {N.}~\bibnamefont {Sueishi}},\ and\ \bibinfo {author} {\bibfnamefont {M.}~\bibnamefont {\"Unsal}},\ }\bibfield  {title} {\bibinfo {title} {{Exact WKB analysis for SUSY and quantum deformed potentials: Quantum mechanics with Grassmann fields and Wess-Zumino terms}},\ }\href {https://doi.org/10.1103/PhysRevD.107.045019} {\bibfield  {journal} {\bibinfo  {journal} {Phys. Rev. D}\ }\textbf {\bibinfo {volume} {107}},\ \bibinfo {pages} {045019} (\bibinfo {year} {2023})},\ \Eprint {https://arxiv.org/abs/2111.05922} {arXiv:2111.05922 [hep-th]} \BibitemShut {NoStop}%
\bibitem [{\citenamefont {Dunne}\ and\ \citenamefont {\"Unsal}(2014{\natexlab{a}})}]{Dunne:2013ada}%
  \BibitemOpen
  \bibfield  {author} {\bibinfo {author} {\bibfnamefont {G.~V.}\ \bibnamefont {Dunne}}\ and\ \bibinfo {author} {\bibfnamefont {M.}~\bibnamefont {\"Unsal}},\ }\bibfield  {title} {\bibinfo {title} {{Generating nonperturbative physics from perturbation theory}},\ }\href {https://doi.org/10.1103/PhysRevD.89.041701} {\bibfield  {journal} {\bibinfo  {journal} {Phys. Rev. D}\ }\textbf {\bibinfo {volume} {89}},\ \bibinfo {pages} {041701} (\bibinfo {year} {2014}{\natexlab{a}})},\ \Eprint {https://arxiv.org/abs/1306.4405} {arXiv:1306.4405 [hep-th]} \BibitemShut {NoStop}%
\bibitem [{\citenamefont {\'{A}lvarez}(2004)}]{Alvarez2004}%
  \BibitemOpen
  \bibfield  {author} {\bibinfo {author} {\bibfnamefont {G.}~\bibnamefont {\'{A}lvarez}},\ }\bibfield  {title} {\bibinfo {title} {{Langer–Cherry derivation of the multi-instanton expansion for the symmetric double well}},\ }\href {https://doi.org/10.1063/1.1767988} {\bibfield  {journal} {\bibinfo  {journal} {J. Math. Phys. (N.Y.)}\ }\textbf {\bibinfo {volume} {45}},\ \bibinfo {pages} {3095} (\bibinfo {year} {2004})}\BibitemShut {NoStop}%
\bibitem [{\citenamefont {Álvarez}\ and\ \citenamefont {Casares}(2000)}]{Alvarez:2000}%
  \BibitemOpen
  \bibfield  {author} {\bibinfo {author} {\bibfnamefont {G.}~\bibnamefont {Álvarez}}\ and\ \bibinfo {author} {\bibfnamefont {C.}~\bibnamefont {Casares}},\ }\bibfield  {title} {\bibinfo {title} {Exponentially small corrections in the asymptotic expansion of the eigenvalues of the cubic anharmonic oscillator},\ }\href {https://doi.org/10.1088/0305-4470/33/29/302} {\bibfield  {journal} {\bibinfo  {journal} {J. Phys. A}\ }\textbf {\bibinfo {volume} {33}},\ \bibinfo {pages} {5171} (\bibinfo {year} {2000})}\BibitemShut {NoStop}%
\bibitem [{\citenamefont {Álvarez}\ \emph {et~al.}(2002)\citenamefont {Álvarez}, \citenamefont {Howls},\ and\ \citenamefont {Silverstone}}]{Alvarez:2002}%
  \BibitemOpen
  \bibfield  {author} {\bibinfo {author} {\bibfnamefont {G.}~\bibnamefont {Álvarez}}, \bibinfo {author} {\bibfnamefont {C.~J.}\ \bibnamefont {Howls}},\ and\ \bibinfo {author} {\bibfnamefont {H.~J.}\ \bibnamefont {Silverstone}},\ }\bibfield  {title} {\bibinfo {title} {Anharmonic oscillator discontinuity formulae up to second-exponentially-small order},\ }\href {https://doi.org/10.1088/0305-4470/35/18/302} {\bibfield  {journal} {\bibinfo  {journal} {J. Phys. A}\ }\textbf {\bibinfo {volume} {35}},\ \bibinfo {pages} {4003} (\bibinfo {year} {2002})}\BibitemShut {NoStop}%
\bibitem [{\citenamefont {Gahramanov}\ and\ \citenamefont {Tezgin}(2016)}]{Gahramanov:2015yxk}%
  \BibitemOpen
  \bibfield  {author} {\bibinfo {author} {\bibfnamefont {I.}~\bibnamefont {Gahramanov}}\ and\ \bibinfo {author} {\bibfnamefont {K.}~\bibnamefont {Tezgin}},\ }\bibfield  {title} {\bibinfo {title} {{Remark on the Dunne-\"Unsal relation in exact semiclassics}},\ }\href {https://doi.org/10.1103/PhysRevD.93.065037} {\bibfield  {journal} {\bibinfo  {journal} {Phys. Rev. D}\ }\textbf {\bibinfo {volume} {93}},\ \bibinfo {pages} {065037} (\bibinfo {year} {2016})},\ \Eprint {https://arxiv.org/abs/1512.08466} {arXiv:1512.08466 [hep-th]} \BibitemShut {NoStop}%
\bibitem [{\citenamefont {Dunne}\ and\ \citenamefont {\"Unsal}(2014{\natexlab{b}})}]{Dunne:2014bca}%
  \BibitemOpen
  \bibfield  {author} {\bibinfo {author} {\bibfnamefont {G.~V.}\ \bibnamefont {Dunne}}\ and\ \bibinfo {author} {\bibfnamefont {M.}~\bibnamefont {\"Unsal}},\ }\bibfield  {title} {\bibinfo {title} {{Uniform WKB, multi-instantons, and resurgent trans-series}},\ }\href {https://doi.org/10.1103/PhysRevD.89.105009} {\bibfield  {journal} {\bibinfo  {journal} {Phys. Rev. D}\ }\textbf {\bibinfo {volume} {89}},\ \bibinfo {pages} {105009} (\bibinfo {year} {2014}{\natexlab{b}})},\ \Eprint {https://arxiv.org/abs/1401.5202} {arXiv:1401.5202 [hep-th]} \BibitemShut {NoStop}%
\bibitem [{\citenamefont {Ba\c{s}ar}\ \emph {et~al.}(2013)\citenamefont {Ba\c{s}ar}, \citenamefont {Dunne},\ and\ \citenamefont {\"Unsal}}]{Basar:2013eka}%
  \BibitemOpen
  \bibfield  {author} {\bibinfo {author} {\bibfnamefont {G.}~\bibnamefont {Ba\c{s}ar}}, \bibinfo {author} {\bibfnamefont {G.~V.}\ \bibnamefont {Dunne}},\ and\ \bibinfo {author} {\bibfnamefont {M.}~\bibnamefont {\"Unsal}},\ }\bibfield  {title} {\bibinfo {title} {{Resurgence theory, ghost-instantons, and analytic continuation of path integrals}},\ }\href {https://doi.org/10.1007/JHEP10(2013)041} {\bibfield  {journal} {\bibinfo  {journal} {J. High Energy Phys.}\ }\textbf {\bibinfo {volume} {2013}}\bibfield  {number} {\bibinfo  {number} { (10)},\ \bibinfo {pages} {041}},\ }\Eprint {https://arxiv.org/abs/1308.1108} {arXiv:1308.1108 [hep-th]} \BibitemShut {NoStop}%
\bibitem [{\citenamefont {Cherman}\ \emph {et~al.}(2015)\citenamefont {Cherman}, \citenamefont {Dorigoni},\ and\ \citenamefont {\"Unsal}}]{Cherman:2014ofa}%
  \BibitemOpen
  \bibfield  {author} {\bibinfo {author} {\bibfnamefont {A.}~\bibnamefont {Cherman}}, \bibinfo {author} {\bibfnamefont {D.}~\bibnamefont {Dorigoni}},\ and\ \bibinfo {author} {\bibfnamefont {M.}~\bibnamefont {\"Unsal}},\ }\bibfield  {title} {\bibinfo {title} {{Decoding perturbation theory using resurgence: Stokes phenomena, new saddle points and Lefschetz thimbles}},\ }\href {https://doi.org/10.1007/JHEP10(2015)056} {\bibfield  {journal} {\bibinfo  {journal} {J. High Energy Phys.}\ }\textbf {\bibinfo {volume} {2015}}\bibfield  {number} {\bibinfo  {number} { (10)},\ \bibinfo {pages} {056}},\ }\Eprint {https://arxiv.org/abs/1403.1277} {arXiv:1403.1277 [hep-th]} \BibitemShut {NoStop}%
\bibitem [{\citenamefont {Behtash}\ \emph {et~al.}(2017)\citenamefont {Behtash}, \citenamefont {Dunne}, \citenamefont {Sch\"afer}, \citenamefont {Sulejmanpasic},\ and\ \citenamefont {\"Unsal}}]{Behtash:2015loa}%
  \BibitemOpen
  \bibfield  {author} {\bibinfo {author} {\bibfnamefont {A.}~\bibnamefont {Behtash}}, \bibinfo {author} {\bibfnamefont {G.~V.}\ \bibnamefont {Dunne}}, \bibinfo {author} {\bibfnamefont {T.}~\bibnamefont {Sch\"afer}}, \bibinfo {author} {\bibfnamefont {T.}~\bibnamefont {Sulejmanpasic}},\ and\ \bibinfo {author} {\bibfnamefont {M.}~\bibnamefont {\"Unsal}},\ }\bibfield  {title} {\bibinfo {title} {{Toward Picard\textendash{}Lefschetz theory of path integrals, complex saddles and resurgence}},\ }\href {https://doi.org/10.4310/AMSA.2017.v2.n1.a3} {\bibfield  {journal} {\bibinfo  {journal} {Ann. Math. Sci. Appl.}\ }\textbf {\bibinfo {volume} {02}},\ \bibinfo {pages} {95} (\bibinfo {year} {2017})},\ \Eprint {https://arxiv.org/abs/1510.03435} {arXiv:1510.03435 [hep-th]} \BibitemShut {NoStop}%
\bibitem [{\citenamefont {Behtash}\ \emph {et~al.}(2016)\citenamefont {Behtash}, \citenamefont {Dunne}, \citenamefont {Sch\"afer}, \citenamefont {Sulejmanpasic},\ and\ \citenamefont {\"Unsal}}]{Behtash:2015zha}%
  \BibitemOpen
  \bibfield  {author} {\bibinfo {author} {\bibfnamefont {A.}~\bibnamefont {Behtash}}, \bibinfo {author} {\bibfnamefont {G.~V.}\ \bibnamefont {Dunne}}, \bibinfo {author} {\bibfnamefont {T.}~\bibnamefont {Sch\"afer}}, \bibinfo {author} {\bibfnamefont {T.}~\bibnamefont {Sulejmanpasic}},\ and\ \bibinfo {author} {\bibfnamefont {M.}~\bibnamefont {\"Unsal}},\ }\bibfield  {title} {\bibinfo {title} {{Complexified path integrals, exact saddles and supersymmetry}},\ }\href {https://doi.org/10.1103/PhysRevLett.116.011601} {\bibfield  {journal} {\bibinfo  {journal} {Phys. Rev. Lett.}\ }\textbf {\bibinfo {volume} {116}},\ \bibinfo {pages} {011601} (\bibinfo {year} {2016})},\ \Eprint {https://arxiv.org/abs/1510.00978} {arXiv:1510.00978 [hep-th]} \BibitemShut {NoStop}%
\bibitem [{\citenamefont {Gahramanov}\ and\ \citenamefont {Tezgin}(2017)}]{Gahramanov:2016xjj}%
  \BibitemOpen
  \bibfield  {author} {\bibinfo {author} {\bibfnamefont {I.}~\bibnamefont {Gahramanov}}\ and\ \bibinfo {author} {\bibfnamefont {K.}~\bibnamefont {Tezgin}},\ }\bibfield  {title} {\bibinfo {title} {{A resurgence analysis for cubic and quartic anharmonic potentials}},\ }\href {https://doi.org/10.1142/S0217751X17500336} {\bibfield  {journal} {\bibinfo  {journal} {Int. J. Mod. Phys. A}\ }\textbf {\bibinfo {volume} {32}},\ \bibinfo {pages} {1750033} (\bibinfo {year} {2017})},\ \Eprint {https://arxiv.org/abs/1608.08119} {arXiv:1608.08119 [hep-th]} \BibitemShut {NoStop}%
\bibitem [{\citenamefont {Ba\c{s}ar}\ and\ \citenamefont {Dunne}(2015)}]{Basar:2015xna}%
  \BibitemOpen
  \bibfield  {author} {\bibinfo {author} {\bibfnamefont {G.}~\bibnamefont {Ba\c{s}ar}}\ and\ \bibinfo {author} {\bibfnamefont {G.~V.}\ \bibnamefont {Dunne}},\ }\bibfield  {title} {\bibinfo {title} {{Resurgence and the Nekrasov-Shatashvili limit: Connecting weak and strong coupling in the Mathieu and Lam\'e systems}},\ }\href {https://doi.org/10.1007/JHEP02(2015)160} {\bibfield  {journal} {\bibinfo  {journal} {J. High Energy Phys.}\ }\textbf {\bibinfo {volume} {2015}}\bibfield  {number} {\bibinfo  {number} { (02)},\ \bibinfo {pages} {160}},\ }\Eprint {https://arxiv.org/abs/1501.05671} {arXiv:1501.05671 [hep-th]} \BibitemShut {NoStop}%
\bibitem [{\citenamefont {Dunne}\ and\ \citenamefont {{\"U}nsal}(2017)}]{Dunne:2016qix}%
  \BibitemOpen
  \bibfield  {author} {\bibinfo {author} {\bibfnamefont {G.~V.}\ \bibnamefont {Dunne}}\ and\ \bibinfo {author} {\bibfnamefont {M.}~\bibnamefont {{\"U}nsal}},\ }\bibfield  {title} {\bibinfo {title} {{WKB and resurgence in the Mathieu equation}},\ }in\ \href {https://doi.org/10.1007/978-88-7642-613-1_6} {\emph {\bibinfo {booktitle} {Resurgence, Physics and Numbers}}},\ \bibinfo {editor} {edited by\ \bibinfo {editor} {\bibfnamefont {F.}~\bibnamefont {Fauvet}}, \bibinfo {editor} {\bibfnamefont {D.}~\bibnamefont {Manchon}}, \bibinfo {editor} {\bibfnamefont {S.}~\bibnamefont {Marmi}},\ and\ \bibinfo {editor} {\bibfnamefont {D.}~\bibnamefont {Sauzin}}}\ (\bibinfo  {publisher} {Scuola Normale Superiore},\ \bibinfo {address} {Pisa},\ \bibinfo {year} {2017})\ pp.\ \bibinfo {pages} {249--298},\ \Eprint {https://arxiv.org/abs/1603.04924} {arXiv:1603.04924 [math-ph]} \BibitemShut {NoStop}%
\bibitem [{\citenamefont {Fujimori}\ \emph {et~al.}(2017{\natexlab{a}})\citenamefont {Fujimori}, \citenamefont {Kamata}, \citenamefont {Misumi}, \citenamefont {Nitta},\ and\ \citenamefont {Sakai}}]{Fujimori:2017osz}%
  \BibitemOpen
  \bibfield  {author} {\bibinfo {author} {\bibfnamefont {T.}~\bibnamefont {Fujimori}}, \bibinfo {author} {\bibfnamefont {S.}~\bibnamefont {Kamata}}, \bibinfo {author} {\bibfnamefont {T.}~\bibnamefont {Misumi}}, \bibinfo {author} {\bibfnamefont {M.}~\bibnamefont {Nitta}},\ and\ \bibinfo {author} {\bibfnamefont {N.}~\bibnamefont {Sakai}},\ }\bibfield  {title} {\bibinfo {title} {{Resurgence structure to all orders of multi-bions in deformed SUSY quantum mechanics}},\ }\href {https://doi.org/10.1093/ptep/ptx101} {\bibfield  {journal} {\bibinfo  {journal} {Prog. Theor. Exp. Phys.}\ }\textbf {\bibinfo {volume} {2017}},\ \bibinfo {pages} {083B02} (\bibinfo {year} {2017}{\natexlab{a}})},\ \Eprint {https://arxiv.org/abs/1705.10483} {arXiv:1705.10483 [hep-th]} \BibitemShut {NoStop}%
\bibitem [{\citenamefont {Fujimori}\ \emph {et~al.}(2019)\citenamefont {Fujimori}, \citenamefont {Kamata}, \citenamefont {Misumi}, \citenamefont {Nitta},\ and\ \citenamefont {Sakai}}]{Fujimori:2018kqp}%
  \BibitemOpen
  \bibfield  {author} {\bibinfo {author} {\bibfnamefont {T.}~\bibnamefont {Fujimori}}, \bibinfo {author} {\bibfnamefont {S.}~\bibnamefont {Kamata}}, \bibinfo {author} {\bibfnamefont {T.}~\bibnamefont {Misumi}}, \bibinfo {author} {\bibfnamefont {M.}~\bibnamefont {Nitta}},\ and\ \bibinfo {author} {\bibfnamefont {N.}~\bibnamefont {Sakai}},\ }\bibfield  {title} {\bibinfo {title} {{Bion non-perturbative contributions versus infrared renormalons in two-dimensional $\mathbb C P^{N-1}$ models}},\ }\href {https://doi.org/10.1007/JHEP02(2019)190} {\bibfield  {journal} {\bibinfo  {journal} {J. High Energy Phys.}\ }\textbf {\bibinfo {volume} {2019}}\bibfield  {number} {\bibinfo  {number} { (02)},\ \bibinfo {pages} {190}},\ }\Eprint {https://arxiv.org/abs/1810.03768} {arXiv:1810.03768 [hep-th]} \BibitemShut {NoStop}%
\bibitem [{\citenamefont {Fujimori}\ \emph {et~al.}(2023)\citenamefont {Fujimori}, \citenamefont {Kamata}, \citenamefont {Misumi}, \citenamefont {Nitta},\ and\ \citenamefont {Sakai}}]{Fujimori:2022lng}%
  \BibitemOpen
  \bibfield  {author} {\bibinfo {author} {\bibfnamefont {T.}~\bibnamefont {Fujimori}}, \bibinfo {author} {\bibfnamefont {S.}~\bibnamefont {Kamata}}, \bibinfo {author} {\bibfnamefont {T.}~\bibnamefont {Misumi}}, \bibinfo {author} {\bibfnamefont {M.}~\bibnamefont {Nitta}},\ and\ \bibinfo {author} {\bibfnamefont {N.}~\bibnamefont {Sakai}},\ }\bibfield  {title} {\bibinfo {title} {{All-order resurgence from complexified path integral in a quantum mechanical system with integrability}},\ }\href {https://doi.org/10.1103/PhysRevD.107.105011} {\bibfield  {journal} {\bibinfo  {journal} {Phys. Rev. D}\ }\textbf {\bibinfo {volume} {107}},\ \bibinfo {pages} {105011} (\bibinfo {year} {2023})},\ \Eprint {https://arxiv.org/abs/2205.07436} {arXiv:2205.07436 [hep-th]} \BibitemShut {NoStop}%
\bibitem [{\citenamefont {Codesido}\ and\ \citenamefont {Mari\~{n}o}(2018)}]{Codesido:2017dns}%
  \BibitemOpen
  \bibfield  {author} {\bibinfo {author} {\bibfnamefont {S.}~\bibnamefont {Codesido}}\ and\ \bibinfo {author} {\bibfnamefont {M.}~\bibnamefont {Mari\~{n}o}},\ }\bibfield  {title} {\bibinfo {title} {{Holomorphic anomaly and quantum mechanics}},\ }\href {https://doi.org/10.1088/1751-8121/aa9e77} {\bibfield  {journal} {\bibinfo  {journal} {J. Phys. A}\ }\textbf {\bibinfo {volume} {51}},\ \bibinfo {pages} {055402} (\bibinfo {year} {2018})},\ \Eprint {https://arxiv.org/abs/1612.07687} {arXiv:1612.07687 [hep-th]} \BibitemShut {NoStop}%
\bibitem [{\citenamefont {Ba\c{s}ar}\ \emph {et~al.}(2017)\citenamefont {Ba\c{s}ar}, \citenamefont {Dunne},\ and\ \citenamefont {\"Unsal}}]{Basar:2017hpr}%
  \BibitemOpen
  \bibfield  {author} {\bibinfo {author} {\bibfnamefont {G.}~\bibnamefont {Ba\c{s}ar}}, \bibinfo {author} {\bibfnamefont {G.~V.}\ \bibnamefont {Dunne}},\ and\ \bibinfo {author} {\bibfnamefont {M.}~\bibnamefont {\"Unsal}},\ }\bibfield  {title} {\bibinfo {title} {{Quantum geometry of resurgent perturbative/nonperturbative relations}},\ }\href {https://doi.org/10.1007/JHEP05(2017)087} {\bibfield  {journal} {\bibinfo  {journal} {J. High Energy Phys.}\ }\textbf {\bibinfo {volume} {2017}}\bibfield  {number} {\bibinfo  {number} { (05)},\ \bibinfo {pages} {087}},\ }\Eprint {https://arxiv.org/abs/1701.06572} {arXiv:1701.06572 [hep-th]} \BibitemShut {NoStop}%
\bibitem [{\citenamefont {Gorsky}\ and\ \citenamefont {Milekhin}(2015)}]{Gorsky:2014lia}%
  \BibitemOpen
  \bibfield  {author} {\bibinfo {author} {\bibfnamefont {A.}~\bibnamefont {Gorsky}}\ and\ \bibinfo {author} {\bibfnamefont {A.}~\bibnamefont {Milekhin}},\ }\bibfield  {title} {\bibinfo {title} {{RG-Whitham dynamics and complex Hamiltonian systems}},\ }\href {https://doi.org/10.1016/j.nuclphysb.2015.03.028} {\bibfield  {journal} {\bibinfo  {journal} {Nucl. Phys. B}\ }\textbf {\bibinfo {volume} {895}},\ \bibinfo {pages} {33} (\bibinfo {year} {2015})},\ \Eprint {https://arxiv.org/abs/1408.0425} {arXiv:1408.0425 [hep-th]} \BibitemShut {NoStop}%
\bibitem [{\citenamefont {Misumi}\ \emph {et~al.}(2015)\citenamefont {Misumi}, \citenamefont {Nitta},\ and\ \citenamefont {Sakai}}]{Misumi:2015dua}%
  \BibitemOpen
  \bibfield  {author} {\bibinfo {author} {\bibfnamefont {T.}~\bibnamefont {Misumi}}, \bibinfo {author} {\bibfnamefont {M.}~\bibnamefont {Nitta}},\ and\ \bibinfo {author} {\bibfnamefont {N.}~\bibnamefont {Sakai}},\ }\bibfield  {title} {\bibinfo {title} {{Resurgence in sine-Gordon quantum mechanics: Exact agreement between multi-instantons and uniform WKB}},\ }\href {https://doi.org/10.1007/JHEP09(2015)157} {\bibfield  {journal} {\bibinfo  {journal} {J. High Energy Phys.}\ }\textbf {\bibinfo {volume} {2015}}\bibfield  {number} {\bibinfo  {number} { (09)},\ \bibinfo {pages} {157}},\ }\Eprint {https://arxiv.org/abs/1507.00408} {arXiv:1507.00408 [hep-th]} \BibitemShut {NoStop}%
\bibitem [{\citenamefont {van Spaendonck}\ and\ \citenamefont {Vonk}(2024)}]{vanSpaendonck:2023znn}%
  \BibitemOpen
  \bibfield  {author} {\bibinfo {author} {\bibfnamefont {A.}~\bibnamefont {van Spaendonck}}\ and\ \bibinfo {author} {\bibfnamefont {M.}~\bibnamefont {Vonk}},\ }\bibfield  {title} {\bibinfo {title} {{Exact instanton transseries for quantum mechanics}},\ }\href {https://doi.org/10.21468/SciPostPhys.16.4.103} {\bibfield  {journal} {\bibinfo  {journal} {SciPost Phys.}\ }\textbf {\bibinfo {volume} {16}},\ \bibinfo {pages} {103} (\bibinfo {year} {2024})},\ \Eprint {https://arxiv.org/abs/2309.05700} {arXiv:2309.05700 [hep-th]} \BibitemShut {NoStop}%
\bibitem [{\citenamefont {Costin}(2008)}]{Costin:2008}%
  \BibitemOpen
  \bibfield  {author} {\bibinfo {author} {\bibfnamefont {O.}~\bibnamefont {Costin}},\ }\href {https://doi.org/10.1201/9781420070323} {\emph {\bibinfo {title} {{Asymptotics and Borel Summability}}}}\ (\bibinfo  {publisher} {CRC Press},\ \bibinfo {address} {New York},\ \bibinfo {year} {2008})\BibitemShut {NoStop}%
\bibitem [{\citenamefont {Dorigoni}(2019)}]{Dorigoni:2014hea}%
  \BibitemOpen
  \bibfield  {author} {\bibinfo {author} {\bibfnamefont {D.}~\bibnamefont {Dorigoni}},\ }\bibfield  {title} {\bibinfo {title} {{An introduction to resurgence, trans-series and alien calculus}},\ }\href {https://doi.org/10.1016/j.aop.2019.167914} {\bibfield  {journal} {\bibinfo  {journal} {Ann. Phys. (N.Y.)}\ }\textbf {\bibinfo {volume} {409}},\ \bibinfo {pages} {167914} (\bibinfo {year} {2019})},\ \Eprint {https://arxiv.org/abs/1411.3585} {arXiv:1411.3585 [hep-th]} \BibitemShut {NoStop}%
\bibitem [{\citenamefont {Mari\~no}(2015)}]{Marino:2015yie}%
  \BibitemOpen
  \bibfield  {author} {\bibinfo {author} {\bibfnamefont {M.}~\bibnamefont {Mari\~no}},\ }\href {https://doi.org/10.1017/CBO9781107705968} {\emph {\bibinfo {title} {{Instantons and Large N: An Introduction to Non-Perturbative Methods in Quantum Field Theory}}}}\ (\bibinfo  {publisher} {Cambridge University Press},\ \bibinfo {address} {Cambridge, England},\ \bibinfo {year} {2015})\BibitemShut {NoStop}%
\bibitem [{\citenamefont {Aniceto}\ \emph {et~al.}(2019)\citenamefont {Aniceto}, \citenamefont {Ba\c{s}ar},\ and\ \citenamefont {Schiappa}}]{Aniceto:2018bis}%
  \BibitemOpen
  \bibfield  {author} {\bibinfo {author} {\bibfnamefont {I.}~\bibnamefont {Aniceto}}, \bibinfo {author} {\bibfnamefont {G.}~\bibnamefont {Ba\c{s}ar}},\ and\ \bibinfo {author} {\bibfnamefont {R.}~\bibnamefont {Schiappa}},\ }\bibfield  {title} {\bibinfo {title} {{A primer on resurgent transseries and their asymptotics}},\ }\href {https://doi.org/10.1016/j.physrep.2019.02.003} {\bibfield  {journal} {\bibinfo  {journal} {Phys. Rep.}\ }\textbf {\bibinfo {volume} {809}},\ \bibinfo {pages} {1} (\bibinfo {year} {2019})},\ \Eprint {https://arxiv.org/abs/1802.10441} {arXiv:1802.10441 [hep-th]} \BibitemShut {NoStop}%
\bibitem [{\citenamefont {Kreshchuk}\ and\ \citenamefont {Gulden}(2019)}]{Gulden_2019}%
  \BibitemOpen
  \bibfield  {author} {\bibinfo {author} {\bibfnamefont {M.}~\bibnamefont {Kreshchuk}}\ and\ \bibinfo {author} {\bibfnamefont {T.}~\bibnamefont {Gulden}},\ }\bibfield  {title} {\bibinfo {title} {{The Picard\textendash{}Fuchs equation in classical and quantum physics: Application to higher-order WKB method}},\ }\href {https://doi.org/10.1088/1751-8121/aaf272} {\bibfield  {journal} {\bibinfo  {journal} {J. Phys. A}\ }\textbf {\bibinfo {volume} {52}},\ \bibinfo {pages} {155301} (\bibinfo {year} {2019})},\ \Eprint {https://arxiv.org/abs/1803.07566} {arXiv:1803.07566 [hep-th]} \BibitemShut {NoStop}%
\bibitem [{\citenamefont {Mironov}\ and\ \citenamefont {Morozov}(2010)}]{Mironov:2009uv}%
  \BibitemOpen
  \bibfield  {author} {\bibinfo {author} {\bibfnamefont {A.}~\bibnamefont {Mironov}}\ and\ \bibinfo {author} {\bibfnamefont {A.}~\bibnamefont {Morozov}},\ }\bibfield  {title} {\bibinfo {title} {{Nekrasov functions and exact Bohr-Sommerfeld integrals}},\ }\href {https://doi.org/10.1007/JHEP04(2010)040} {\bibfield  {journal} {\bibinfo  {journal} {J. High Energy Phys.}\ }\textbf {\bibinfo {volume} {2010}}\bibfield  {number} {\bibinfo  {number} { (04)},\ \bibinfo {pages} {040}},\ }\Eprint {https://arxiv.org/abs/0910.5670} {arXiv:0910.5670 [hep-th]} \BibitemShut {NoStop}%
\bibitem [{\citenamefont {Fischbach}\ \emph {et~al.}(2019)\citenamefont {Fischbach}, \citenamefont {Klemm},\ and\ \citenamefont {Nega}}]{Klemm_2019}%
  \BibitemOpen
  \bibfield  {author} {\bibinfo {author} {\bibfnamefont {F.}~\bibnamefont {Fischbach}}, \bibinfo {author} {\bibfnamefont {A.}~\bibnamefont {Klemm}},\ and\ \bibinfo {author} {\bibfnamefont {C.}~\bibnamefont {Nega}},\ }\bibfield  {title} {\bibinfo {title} {{WKB method and quantum periods beyond genus one}},\ }\href {https://doi.org/10.1088/1751-8121/aae8b0} {\bibfield  {journal} {\bibinfo  {journal} {J. Phys. A}\ }\textbf {\bibinfo {volume} {52}},\ \bibinfo {pages} {075402} (\bibinfo {year} {2019})},\ \Eprint {https://arxiv.org/abs/1803.11222} {arXiv:1803.11222 [hep-th]} \BibitemShut {NoStop}%
\bibitem [{\citenamefont {Bender}\ and\ \citenamefont {Wu}(1969)}]{Bender:1969si}%
  \BibitemOpen
  \bibfield  {author} {\bibinfo {author} {\bibfnamefont {C.~M.}\ \bibnamefont {Bender}}\ and\ \bibinfo {author} {\bibfnamefont {T.~T.}\ \bibnamefont {Wu}},\ }\bibfield  {title} {\bibinfo {title} {{Anharmonic oscillator}},\ }\href {https://doi.org/10.1103/PhysRev.184.1231} {\bibfield  {journal} {\bibinfo  {journal} {Phys. Rev.}\ }\textbf {\bibinfo {volume} {184}},\ \bibinfo {pages} {1231} (\bibinfo {year} {1969})}\BibitemShut {NoStop}%
\bibitem [{\citenamefont {Bender}\ and\ \citenamefont {Wu}(1973)}]{Bender:1973rz}%
  \BibitemOpen
  \bibfield  {author} {\bibinfo {author} {\bibfnamefont {C.~M.}\ \bibnamefont {Bender}}\ and\ \bibinfo {author} {\bibfnamefont {T.~T.}\ \bibnamefont {Wu}},\ }\bibfield  {title} {\bibinfo {title} {{Anharmonic oscillator. II. A study of perturbation theory in large order}},\ }\href {https://doi.org/10.1103/PhysRevD.7.1620} {\bibfield  {journal} {\bibinfo  {journal} {Phys. Rev. D}\ }\textbf {\bibinfo {volume} {7}},\ \bibinfo {pages} {1620} (\bibinfo {year} {1973})}\BibitemShut {NoStop}%
\bibitem [{\citenamefont {Sulejmanpasic}\ and\ \citenamefont {\"Unsal}(2018)}]{Sulejmanpasic:2016fwr}%
  \BibitemOpen
  \bibfield  {author} {\bibinfo {author} {\bibfnamefont {T.}~\bibnamefont {Sulejmanpasic}}\ and\ \bibinfo {author} {\bibfnamefont {M.}~\bibnamefont {\"Unsal}},\ }\bibfield  {title} {\bibinfo {title} {{Aspects of perturbation theory in quantum mechanics: The BenderWu Mathematica \textregistered{} package}},\ }\href {https://doi.org/10.1016/j.cpc.2017.11.018} {\bibfield  {journal} {\bibinfo  {journal} {Comput. Phys. Commun.}\ }\textbf {\bibinfo {volume} {228}},\ \bibinfo {pages} {273} (\bibinfo {year} {2018})},\ \Eprint {https://arxiv.org/abs/1608.08256} {arXiv:1608.08256 [hep-th]} \BibitemShut {NoStop}%
\bibitem [{\citenamefont {Fujimori}\ \emph {et~al.}(2016)\citenamefont {Fujimori}, \citenamefont {Kamata}, \citenamefont {Misumi}, \citenamefont {Nitta},\ and\ \citenamefont {Sakai}}]{Fujimori:2016ljw}%
  \BibitemOpen
  \bibfield  {author} {\bibinfo {author} {\bibfnamefont {T.}~\bibnamefont {Fujimori}}, \bibinfo {author} {\bibfnamefont {S.}~\bibnamefont {Kamata}}, \bibinfo {author} {\bibfnamefont {T.}~\bibnamefont {Misumi}}, \bibinfo {author} {\bibfnamefont {M.}~\bibnamefont {Nitta}},\ and\ \bibinfo {author} {\bibfnamefont {N.}~\bibnamefont {Sakai}},\ }\bibfield  {title} {\bibinfo {title} {{Nonperturbative contributions from complexified solutions in $\mathbb{C}P^{N-1}$models}},\ }\href {https://doi.org/10.1103/PhysRevD.94.105002} {\bibfield  {journal} {\bibinfo  {journal} {Phys. Rev. D}\ }\textbf {\bibinfo {volume} {94}},\ \bibinfo {pages} {105002} (\bibinfo {year} {2016})},\ \Eprint {https://arxiv.org/abs/1607.04205} {arXiv:1607.04205 [hep-th]} \BibitemShut {NoStop}%
\bibitem [{\citenamefont {Fujimori}\ \emph {et~al.}(2017{\natexlab{b}})\citenamefont {Fujimori}, \citenamefont {Kamata}, \citenamefont {Misumi}, \citenamefont {Nitta},\ and\ \citenamefont {Sakai}}]{Fujimori:2017oab}%
  \BibitemOpen
  \bibfield  {author} {\bibinfo {author} {\bibfnamefont {T.}~\bibnamefont {Fujimori}}, \bibinfo {author} {\bibfnamefont {S.}~\bibnamefont {Kamata}}, \bibinfo {author} {\bibfnamefont {T.}~\bibnamefont {Misumi}}, \bibinfo {author} {\bibfnamefont {M.}~\bibnamefont {Nitta}},\ and\ \bibinfo {author} {\bibfnamefont {N.}~\bibnamefont {Sakai}},\ }\bibfield  {title} {\bibinfo {title} {{Exact resurgent trans-series and multibion contributions to all orders}},\ }\href {https://doi.org/10.1103/PhysRevD.95.105001} {\bibfield  {journal} {\bibinfo  {journal} {Phys. Rev. D}\ }\textbf {\bibinfo {volume} {95}},\ \bibinfo {pages} {105001} (\bibinfo {year} {2017}{\natexlab{b}})},\ \Eprint {https://arxiv.org/abs/1702.00589} {arXiv:1702.00589 [hep-th]} \BibitemShut {NoStop}%
\bibitem [{\citenamefont {Dunne}\ and\ \citenamefont {\"Unsal}(2016)}]{Dunne:2016jsr}%
  \BibitemOpen
  \bibfield  {author} {\bibinfo {author} {\bibfnamefont {G.~V.}\ \bibnamefont {Dunne}}\ and\ \bibinfo {author} {\bibfnamefont {M.}~\bibnamefont {\"Unsal}},\ }\bibfield  {title} {\bibinfo {title} {{Deconstructing zero: resurgence, supersymmetry and complex saddles}},\ }\href {https://doi.org/10.1007/JHEP12(2016)002} {\bibfield  {journal} {\bibinfo  {journal} {J. High Energy Phys.}\ }\textbf {\bibinfo {volume} {2016}}\bibfield  {number} {\bibinfo  {number} { (12)},\ \bibinfo {pages} {002}},\ }\Eprint {https://arxiv.org/abs/1609.05770} {arXiv:1609.05770 [hep-th]} \BibitemShut {NoStop}%
\bibitem [{\citenamefont {Koz\c{c}az}\ \emph {et~al.}(2018)\citenamefont {Koz\c{c}az}, \citenamefont {Sulejmanpasic}, \citenamefont {Tanizaki},\ and\ \citenamefont {\"Unsal}}]{Kozcaz:2016wvy}%
  \BibitemOpen
  \bibfield  {author} {\bibinfo {author} {\bibfnamefont {C.}~\bibnamefont {Koz\c{c}az}}, \bibinfo {author} {\bibfnamefont {T.}~\bibnamefont {Sulejmanpasic}}, \bibinfo {author} {\bibfnamefont {Y.}~\bibnamefont {Tanizaki}},\ and\ \bibinfo {author} {\bibfnamefont {M.}~\bibnamefont {\"Unsal}},\ }\bibfield  {title} {\bibinfo {title} {{Cheshire Cat resurgence, self-resurgence and quasi-exact solvable systems}},\ }\href {https://doi.org/10.1007/s00220-018-3281-y} {\bibfield  {journal} {\bibinfo  {journal} {Commun. Math. Phys.}\ }\textbf {\bibinfo {volume} {364}},\ \bibinfo {pages} {835} (\bibinfo {year} {2018})},\ \Eprint {https://arxiv.org/abs/1609.06198} {arXiv:1609.06198 [hep-th]} \BibitemShut {NoStop}%
\bibitem [{\citenamefont {Dorigoni}\ and\ \citenamefont {Glass}(2018)}]{Dorigoni:2017smz}%
  \BibitemOpen
  \bibfield  {author} {\bibinfo {author} {\bibfnamefont {D.}~\bibnamefont {Dorigoni}}\ and\ \bibinfo {author} {\bibfnamefont {P.}~\bibnamefont {Glass}},\ }\bibfield  {title} {\bibinfo {title} {{The grin of Cheshire cat resurgence from supersymmetric localization}},\ }\href {https://doi.org/10.21468/SciPostPhys.4.2.012} {\bibfield  {journal} {\bibinfo  {journal} {SciPost Phys.}\ }\textbf {\bibinfo {volume} {4}},\ \bibinfo {pages} {012} (\bibinfo {year} {2018})},\ \Eprint {https://arxiv.org/abs/1711.04802} {arXiv:1711.04802 [hep-th]} \BibitemShut {NoStop}%
\bibitem [{\citenamefont {Dorigoni}\ and\ \citenamefont {Glass}(2019)}]{Dorigoni:2019kux}%
  \BibitemOpen
  \bibfield  {author} {\bibinfo {author} {\bibfnamefont {D.}~\bibnamefont {Dorigoni}}\ and\ \bibinfo {author} {\bibfnamefont {P.}~\bibnamefont {Glass}},\ }\bibfield  {title} {\bibinfo {title} {{Picard-Lefschetz decomposition and Cheshire Cat resurgence in 3D $ \mathcal{N} $ = 2 field theories}},\ }\href {https://doi.org/10.1007/JHEP12(2019)085} {\bibfield  {journal} {\bibinfo  {journal} {J. High Energy Phys.}\ }\textbf {\bibinfo {volume} {2019}}\bibfield  {number} {\bibinfo  {number} { (12)},\ \bibinfo {pages} {085}},\ }\Eprint {https://arxiv.org/abs/1909.05262} {arXiv:1909.05262 [hep-th]} \BibitemShut {NoStop}%
\bibitem [{\citenamefont {Brezin}\ \emph {et~al.}(1977)\citenamefont {Brezin}, \citenamefont {Le~Guillou},\ and\ \citenamefont {Zinn-Justin}}]{Brezin:1976wa}%
  \BibitemOpen
  \bibfield  {author} {\bibinfo {author} {\bibfnamefont {E.}~\bibnamefont {Brezin}}, \bibinfo {author} {\bibfnamefont {J.~C.}\ \bibnamefont {Le~Guillou}},\ and\ \bibinfo {author} {\bibfnamefont {J.}~\bibnamefont {Zinn-Justin}},\ }\bibfield  {title} {\bibinfo {title} {{Perturbation theory at large order. II. Role of the vacuum instability}},\ }\href {https://doi.org/10.1103/PhysRevD.15.1558} {\bibfield  {journal} {\bibinfo  {journal} {Phys. Rev. D}\ }\textbf {\bibinfo {volume} {15}},\ \bibinfo {pages} {1558} (\bibinfo {year} {1977})}\BibitemShut {NoStop}%
\end{thebibliography}%

\end{document}